\documentclass[acmsmall]{acmart}

\usepackage{graphicx}
\usepackage{caption}

\usepackage{algorithm}
\usepackage[noend]{algpseudocode}
\usepackage{float} 
\usepackage{subfigure}
\usepackage{multirow}
\usepackage{hyperref}

\AtBeginDocument{%
  \providecommand\BibTeX{{%
    \normalfont B\kern-0.5em{\scshape i\kern-0.25em b}\kern-0.8em\TeX}}}

\begin{document}

\title[Fairness-aware Sequential Recommendation]{FairSR: Fairness-aware Sequential Recommendation through Multi-Task Learning with Preference Graph Embeddings}

\author{Cheng-Te Li}
\email{chengte@mail.ncku.edu.tw}
\orcid{0000-0001-7995-4787}
\affiliation{%
  \institution{National Cheng Kung University}
  \streetaddress{No. 1, University Rd., East District}
  \city{Tainan City}
  \country{Taiwan}
  \postcode{701}
}

\author{Cheng Hsu}
\affiliation{%
  \institution{National Cheng Kung University}
  \streetaddress{No. 1, University Rd., East District}
  \city{Tainan City}
  \country{Taiwan}
  \postcode{701}
  }
\email{hsuchengmath@gmail.com}

\author{Yang Zhang}
\affiliation{%
  \institution{CISPA Helmholtz Center for Information Security}
  \streetaddress{Stuhlsatzenhaus 5}
  \city{Saarbrücken}
  \country{Germany}
  \postcode{66123}
}
\email{zhang@cispa.de}

\renewcommand{\shortauthors}{C.-T. Li et al.}

\begin{abstract}
Sequential recommendation (SR) learns from the temporal dynamics of user-item interactions to predict the next ones. Fairness-aware recommendation mitigates a variety of algorithmic biases in the learning of user preferences. This paper aims at bringing a marriage between SR and algorithmic fairness. We propose a novel fairness-aware sequential recommendation task, in which a new metric, \textit{interaction fairness}, is defined to estimate how recommended items are fairly interacted by users with different protected attribute groups. We propose a multi-task learning based deep end-to-end model, FairSR, which consists of two parts. One is to learn and distill personalized sequential features from the given user and her item sequence for SR. The other is fairness-aware preference graph embedding (FPGE). The aim of FPGE is two-fold: incorporating the knowledge of users' and items' attributes and their correlation into entity representations, and alleviating the unfair distributions of user attributes on items. Extensive experiments conducted on three datasets show FairSR can outperform state-of-the-art SR models in recommendation performance. In addition, the recommended items by FairSR also exhibit promising interaction fairness.
\end{abstract}


\setcopyright{acmcopyright}
\acmJournal{TIST}
\acmYear{2022} \acmVolume{13} \acmNumber{1} \acmArticle{16} \acmMonth{2} \acmPrice{15.00}\acmDOI{10.1145/3495163}


\maketitle

\section{Introduction}
Sequential recommendation (SR) is a crucial research task in recommender systems (RS)~\cite{recsurv19}. SR considers the chronological order of user-item interactions, and models how users' recent successively accessed items affect the choices of the next ones. To be specific, given a recent sequence of items interacted by a user, SR aims at learning from the sequence to find which items will be interacted by her in the near future. The SR task differs from conventional RS. While RS tends to capture the global user preferences on items~\cite{mfhe16,ngcf19}, SR imposes the sequential dynamics of user-item interactions. Hence, SR requires the learning of long-term and short-term interests and intents of users~\cite{caser18,hgn19} in predicting the next items.

On the other hand, while algorithmic fairness is getting attention in the machine learning community~\cite{fsurv19}, fairness in recommender systems is investigated in various aspects. Typical issues on recommendation fairness can be mainly divided into three groups. The first is dealing with \textit{popularity bias} concerning that few popular items are over-represented in the models~\cite{biasatt18,fbias}. The second is tacking \textit{demographic bias}, in which the representations of users with imbalanced attributes (e.g., gender and age) cannot be equally learned and lead to differentiated and unfair performance~\cite{biasatt18,fairhyb18,ffire19}. The third is imposing \textit{statistical parity} into recommendation, which aims at ensuring similar probability distributions of item ratings for those users in different protected attribute groups~\cite{faircf17,ften18}.

Despite existing methods on fairness-aware recommendation receive satisfactory results, we think there remain several opportunities and challenges. First, to the best of our knowledge, none of the existing studies target at considering fairness into sequential recommendation. This work is an essential attempt to define and solve a fairness-aware SR problem. Second, previous fairness-aware recommendations discussed above are not aware of the \textit{filter bubble} effect~\cite{fbori11}, which states that online personalization tends to effectively isolate users from a diversity of viewpoints. Since people prefer to interact with what they liked or interacted before, learning-based recommenders will reinforce user preferences to satisfy them~\cite{fltbb14}. We regard it as a kind of unfair recommendation for users who want to pursue novel or diverse items. Therefore, it is worthwhile to design a new fairness concept so that recommenders can follow to generate items against the filter bubble. Third, knowledge graphs (KG) that connect items based on their metadata or attributes had been proven to be promising for recommendation systems~\cite{jkgr19,kgat19,mkr19}. Yet KG is not explored for sequential recommendation. Besides, while existing KG models item-item relationships, user attributes can be incorporated into KG so that the item embeddings can encode the knowledge on user traits.

\begin{figure}[!t]
  \centering
  \includegraphics[width=0.95\linewidth]{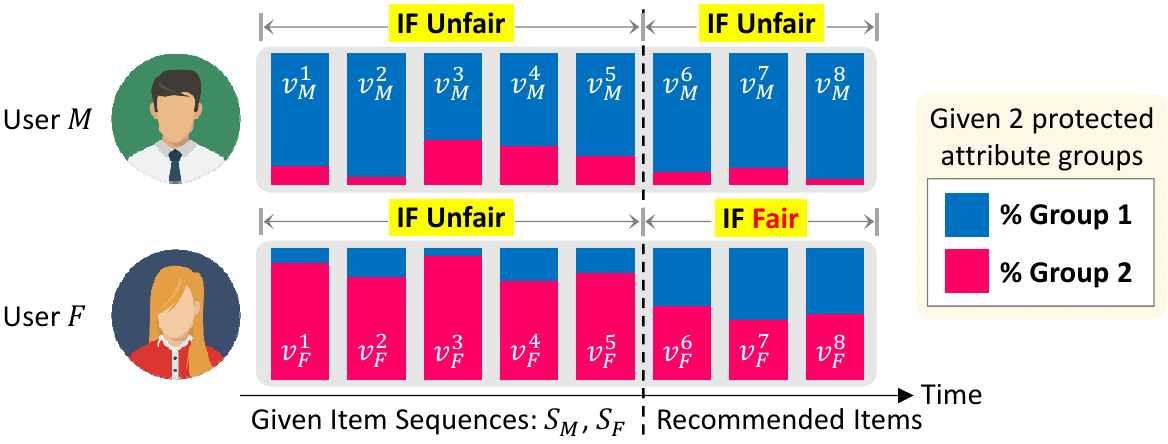}
  \caption{Elaboration of fairness-aware SR. Note that for each user, every colored bar is an item, and their associated notations $v_M^i$ and $v_M^i$ indicate the $i$-th item that user $M$ and $F$ interact with. The Blue and red parts inside each bar exhibit the proportions of historical interactions with an item created by users belonging to attribute groups 1 and 2, respectively.}
  \label{fig:fsr}
\end{figure}

In this paper, we propose a novel fairness-aware sequential recommendation task. Given user-attribute groups to be protected, i.e., a set of attribute groups that concerned for fairness (e.g., ``Female \& Age 20-29'' and ``Male \& Age 50-59''), and the recent item subsequence of a user, we aim to not only accurately predict the next items, but also require that the recommended items lead to \textit{Interaction Fairness} (IF). Better IF means that recommended items tend to be equally interacted by users of different protected attribute groups. Taking Figure~\ref{fig:fsr} as an example, we assume two protected attribute groups 1 and 2 are specified. Given sequences of $5$ items for users M and F, the recommended items of F are fairer in terms of IF, while those of M are unfair. The reason is each recommended item of F tends to be equally interacted by groups 1 and 2, but it is not for M. Note that although the given item sequences of users are not fair, we still require their recommended next ones to be fair in terms of IF. 
When fighting against the filter bubbles, online social media platforms are in need of interaction fairness for their recommender systems.
If an SR system is aware of interaction fairness, a user will have high possibility to receive items being interacted by other users with diverse attribute groups, and the effect of filter bubbles can be mitigated.

As aforementioned, existing studies have investigated three common types of unfairness issues in recommendation, including popularity bias, demographic bias, and statistical parity. The proposed interaction fairness (IF) in SR is to mitigate the filter bubble effect in recommender systems, rather than solving the three types of unfair recommendation. The difference between IF and the three types of unfairness is two-fold. First, IF considers how other users interact with the recommended items, which are expected to be fairly treated by user groups with different protected attributes. Both popularity bias and demographic bias concern about users/items with less attention (i.e., less popular items, the minority attribute of users), and the proposed methods want them to be fairly treated in the construction of recommender systems. Second, the recommended items with higher IF values provide higher diversity for users, but those with maintaining statistical parity is to keep the distribution of item ratings the same before and after recommendation, instead of dealing with the issue of diversity.

We propose a multi-task learning (MTL) based end-to-end deep model, \texttt{FairSR}, to solve the fairness-aware sequential recommendation task. The main task aims to perform sequential recommendation (SR) by learning sequential features from the given item sequence. A personalized feature gating as well as two convolution mechanisms are performed to produce effective sequence representations that encode user preferences and sequential patterns. Another task of FairSR is to learn fairness-aware preference graph embeddings (FPGE). Borrowing from the idea of knowledge graph embeddings~\cite{kgesur17}, we encode both attributes of users and items and their relations into entity embeddings. While user-item interactions are biased to some attribute groups, we propose a fairness-aware triplet sampling to generate positive triplets of the head, relation, and tail so that the bias can be mitigated in FPGE. Two tasks are connected with each other through a cross item-preference learning (CIPL), which encodes shared features between items in SR and their corresponding entities in FPGE.

We summarize the contribution of this paper as follows.
\begin{itemize}
\item We propose a novel fairness-aware sequential recommendation problem that brings a marriage between SR and recommendation fairness. We define a new fairness metric, interaction fairness, which quantifies the degree of the filter bubble effect by estimating how recommended items are interacted by protected attribute groups.
\item We develop a new multi-task learning based deep model, \texttt{FairSR}~\footnote{The FairSR code and datasets can be accessed via this link: \url{https://github.com/fairsr/fairsr}}, to solve the problem. The main task is SR that predicts the next items based on the learned personalized sequential features. The other task is FPGE that encodes both knowledge and fairness of attributes into entity embeddings.
\item Extensive experiments conducted on three real datasets exhibit not only promising recommendation performance of FairSR, compared to several state-of-the-art models, but also fair recommendation in terms of interaction fairness.
\end{itemize}

We organize this paper as follows. Section~\ref{sec-related} reviews relevant studies, followed by Section~\ref{sec-prob} that describes the problem statement. In Section~\ref{sec-model}, we present the technical details of the proposed FairSR model. We give the experimental results in Section~\ref{sec-exp}, and conclude this work in Section~\ref{sec-conclude}.
\section{Related Work}
\label{sec-related}
The relevant studies can be categorized into three groups, sequential recommendation (SR) models, fairness-aware recommender systems (RS), and knowledge graph (KG) enhanced recommender systems.

\textbf{SR Models.} In deep SR models, recurrent neural networks~\cite{gru4rec16,rrn17,rnnsr18} and convolutional neural networks~\cite{caser18} are used to extract long-term and short-term sequential features. SASRec~\cite{sasrec18} is a self-attentive model that can identify the most significant items for prediction. HGN~\cite{hgn19} is a hierarchical gating neural network that adopts feature gating and instance gating to determine what item features should be used for recommendation. MARank~\cite{marank19} is a multi-order attentive ranking model that unifies both individual- and union-level item-item interaction into the preference inference model from multiple views.  NextItNet~\cite{nextitnet19} is a dilated convolution-based generative method to learn long-range dependencies in the item sequence. JODIE~\cite{jodie19} is a coupled recurrent neural network model that jointly learns the embedding trajectories of users and items from a sequence of temporal interactions. Recent studies on SR deal with various limitations in real-world scenarios. RetaGNN~\cite{retagnn21} delivers a graph neural network-based holistic sequential recommendation model that accommodates SR under conventional, inductive, and transferable settings. Mecos~\cite{cmeta21} concentrates on mitigating the item cold-start problem in SR without utilizing side information. A category-aware collaborative SR model~\cite{catesr21} further proposes to improve the performance by utilizing multi-grained category information of items.

\textbf{Fairness-aware RS.} Various kinds of fairness are investigated and developed for RS. The most well-known three are popularity bias~\cite{biasatt18,fbias}, demographic bias~\cite{biasatt18,fairhyb18,ffire19}, and statistical parity: \cite{faircf17,ften18}. Popularity bias means a few popular items are over-represented in the models. A personalized diversification re-ranking approach is developed to increase the representations of unpopular items~\cite{dpbias19}. Demographic bias indicates that the representations of users with imbalanced attributes (e.g., gender and age) cannot be equally learned and lead to differentiated and unfair performance. 
Incorporating multiple-source data~\cite{fairhyb18} and performing data augmentation~\cite{ffire19} can alleviate demographic bias. Statistical parity aims at ensuring similar probability distributions of item ratings for users in different protected attribute groups. Extracting and isolating sensitive information from the factorized matrices~\cite{ften18} and re-ranking recommended items based on required attribute distributions~\cite{frank19} can improve statistical parity. In addition to such three fairness issues, Jiang et al.~\cite{accu18} deal with account-level recommendation bias by identifying users. Although fairness-aware embedding learning methods~\cite{fwalk19,cfgrl19} fairly encode attributes for node representation learning in graphs, they do not target at sequential recommendation. Beutel et al.~\cite{fpwc19} devise a pairwise regularizer to improve pairwise fairness metrics in RS. Rather than considering fairness in RS, our work aims at imposing fairness into SR.


\textbf{KG-enhanced RS.} Knowledge graph (KG) embedding~\cite{kgesur17} provides additional features depicting the association between items through metadata and attributes. KGs are leveraged in various ways, including propagating user preferences over knowledge entities by RippleNet~\cite{rpnet18}, multi-task learning with KG Embedding by MKR~\cite{mkr19}, applying graph attention on a user-item-attribute graph by KGAT~\cite{kgat19}, adopting LSTM to model sequential dependencies of entities and relations~\cite{kprn19}, and integrating induction of explainable rules from KG by RuleRec~\cite{exrule19}. MARINE~\cite{marine19} combines homogeneous and heterogeneous graph embedding learning mechanisms to recommend links between entities.
Furthermore, KGPL~\cite{kgpl21} assigns pseudo-positive labels to unobserved samples through knowledge graph neural network-based predictions so that the recommendation model can better deal with the cold-start issues. KGPolicy~\cite{kgpmf20} leverages rich relations between items in the knowledge graphs to sample high-quality negatives and boost the performance of recommenders. JNSKR~\cite{jnskr20} presents a non-sampling and efficient learning mechanism based on an attentive neural network for better knowledge graph-enhanced recommenders. KGIN~\cite{kgin21} learns user intents by modeling attentive combinations of relations in the knowledge graph to enhance the recommendation performance and bring model interpretability.
Although these studies successfully apply knowledge graphs with various mechanisms for better recommender systems, exploiting KG to assist sequential recommendation is not well explored yet.

The main difference between our work and existing studies consists of three parts. First, in the task of sequential recommendation, none of the existing methods (e.g., \cite{caser18,mkr19,sasrec18,hgn19,rpnet18}) can simultaneously model the personalized sequential features for users and the latent correlation behind items via the concept of knowledge graphs. We believe that item representation learning by utilizing the relations between items and attributes can better model user preferences. Second, while some of the conventional recommendation models are devised to mitigate various unfairness issues (e.g., FATR~\cite{ften18} for statistical parity, Fair-PSL~\cite{biasatt18} for demographic bias, and miscalibration~\cite{fbias} for popularity bias), existing SR methods cannot deal with any fairness issues. The proposed Fairness-aware Triplet Sampling mechanism in FairSR brings the item embeddings to be aware of interaction fairness. Third, in FariSR, we devise the Personalized Feature Gating component, which originates from gated recurrent unit~\cite{glu17} in the task of language modeling, to adaptively select items specialized to the targeted user.

\section{Problem Statement}
\label{sec-prob}
In typical recommendation, we have a set of $M$ users $\mathcal{U}=\{u_1, ..., u_M\}$ and a set of $N$ items $\mathcal{V}=\{v_1,...,v_N\}$. The matrix of user-item interactions is denoted by $\mathbf{Y}\in \mathbb{R}^{M\times N}$ based on the implicit feedback from users, in which $y_{uv}=1$ indicates user $u$ had ever interacted with item $v$; otherwise, $y_{uv}=0$. While a user $u$ sequentially interact with different items in a chronological manner, we denote the corresponding item sequence as $\mathcal{S}^u=(s^u_1, s^u_2, ..., s^u_L)$, where $L=|\mathcal{S}^u|$ and $s^u_i\in V$ is an item index that user $u$ has interacted with. 

Let $A$ be the universal set of demographic attribute groups. A demographic attribute group $a\in A$ can be the gender (\texttt{g}), age (\texttt{o}), country (\texttt{c}), or their combinations. We denote the set of all possible values of $a$ as $Z^a$, and denote a specific value as $z^a\in Z^a$. Let the function $\mathcal{A}^{a}: \mathcal{U}\rightarrow Z^a$ map a user to her attribute groups. An attribute group value $\mathcal{A}^{a}(u)$ of user $u$ can be, for example, ``male'' or ``female'' for attribute group $\{$\texttt{Gender}$\}$, and ``10-19 \& US'', ``20-29 \& UK'' or ``30-39 \& JP'' for attribute group $\{$\texttt{Age}, \texttt{Country}$\}$. To enhance the readability, we use \textit{attribute} and \textit{attribute value} to represent \textit{demographic attribute group} and \textit{demographic attribute group value}, respectively, throughout the paper.

We consider that fair recommendation expects an \textit{unbiased} distribution on recommended items with respect to a certain attribute. Hence, a criterion is required to measure the \textit{equality} of user interactions with items. Given an attribute value $z^a$ of attribute $a\in A$, we first define \textit{adoption proportion} with respect to a certain item.
The adoption proportion measures the percentage of users with a specific attribute value adopting an item among all users who had ever interacted with the item. Higher values of adoption proportion (e.g., $\approx 1.0$) indicate that the item is highly biased to a specific attribute value, and thus tends to be unfair with respect to the attribute.
The definition of adoption proportion is given by: 
\begin{equation}
p_v(z^a)=\frac{|\{ u | \mathcal{A}^{a}(u)=z^a, u\in \mathcal{U}(v) \}|}{|\mathcal{U}(v)|},   
\end{equation}
where $\mathcal{U}(v)$ is the set of users who had ever interacted with item $v$. 

Since an item can be adopted by users with different values of a particular attribute, we further define the \textit{adoption equality} to collectively quantify the degree that an item is equally interacted based on adoption proportion using all values of an attribute. The adoption equality considers all possible values of an attribute to measure the fairness of an item being interacted. If each value of an attribute contributes nearly the same to an item in terms of adoption proportion, the adoption equality gets a higher score, and we can say that this item tends to be fairly interacted by users with different values of that attribute. That said, the adoption of the item is not biased to users with specific attribute values.
Then we define \textit{adoption equality} of item $v$, denoted by $e^a(v)$, based on information entropy, given by: 
\begin{equation}
e^a(v)=-\sum_{z^a\in Z^a} p_v(z^a) \log p_v(z^a).
\end{equation}
A higher value of $e^a(v)$ indicates higher adoption equality of item $v$. 

\begin{figure*}[!t]
  \centering
  \includegraphics[width=1.0\textwidth]{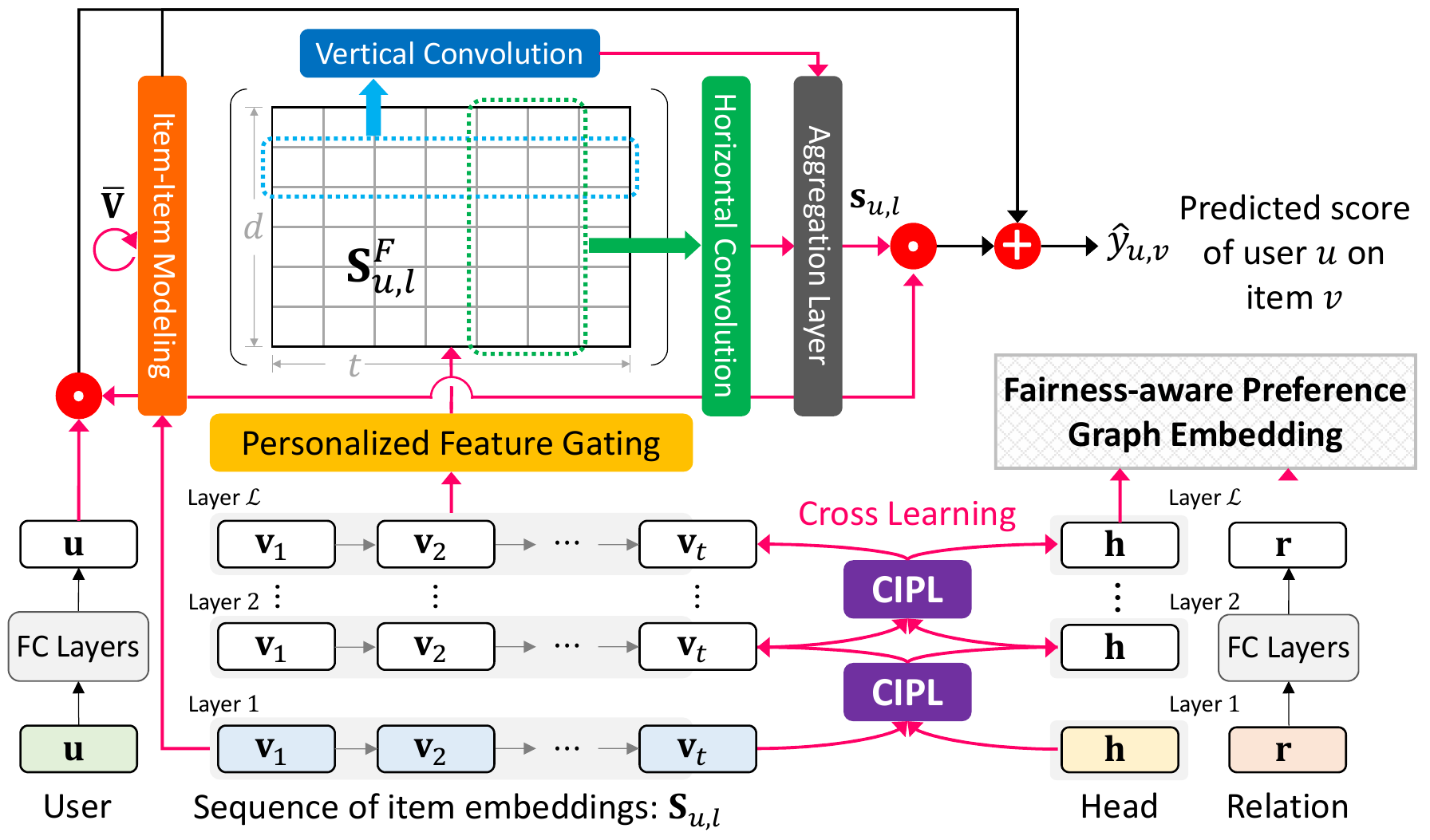}
  \caption{The main framework of FairSR.}
  \label{fig:model}
\end{figure*}

Given an attribute $a\in A$, for a list of items $\mathcal{S}^u_{t:L}$ ($t<L$) that are recommended to user $u$, we define its corresponding \textit{interaction fairness} (IF) score $\mathcal{F}^a(u)$ with respect to attribute $a$ based on adoption equality, given by:
\begin{equation}
\mathcal{F}^a(u)=\sum_{v\in \mathcal{S}^u_{t:L}} e^a(v).
\end{equation}
A higher value of $\mathcal{F}^a(u)$ indicates that a recommendation algorithm tends to recommend items to user $u$ with a higher fairness degree for attribute $a$. That said, the recommended items tend to be unbiased to a specific attribute value $z^a\in Z^a$.

The \textit{fairness-aware sequential recommendation} (FairSR) problem can be defined as below. Given the earlier subsequence $\mathcal{S}^u_{1:t}$ ($t<L$) of every user $u\in \mathcal{U}$, we aim to recommend a list of items from item set $\mathcal{V}$ to each user, i.e., predict whether user $u$ will interact with item $v\in \mathcal{V}$ after time $t$ (whether the items in ground truth $\mathcal{S}^u_{t:L}$ will appear in the recommended list), and maximize the corresponding interaction fairness $\mathcal{F}^a(u)$.

\section{The Proposed FairSR Model}
\label{sec-model}
\begin{figure}[!t]
  \centering
  \includegraphics[width=0.6\linewidth]{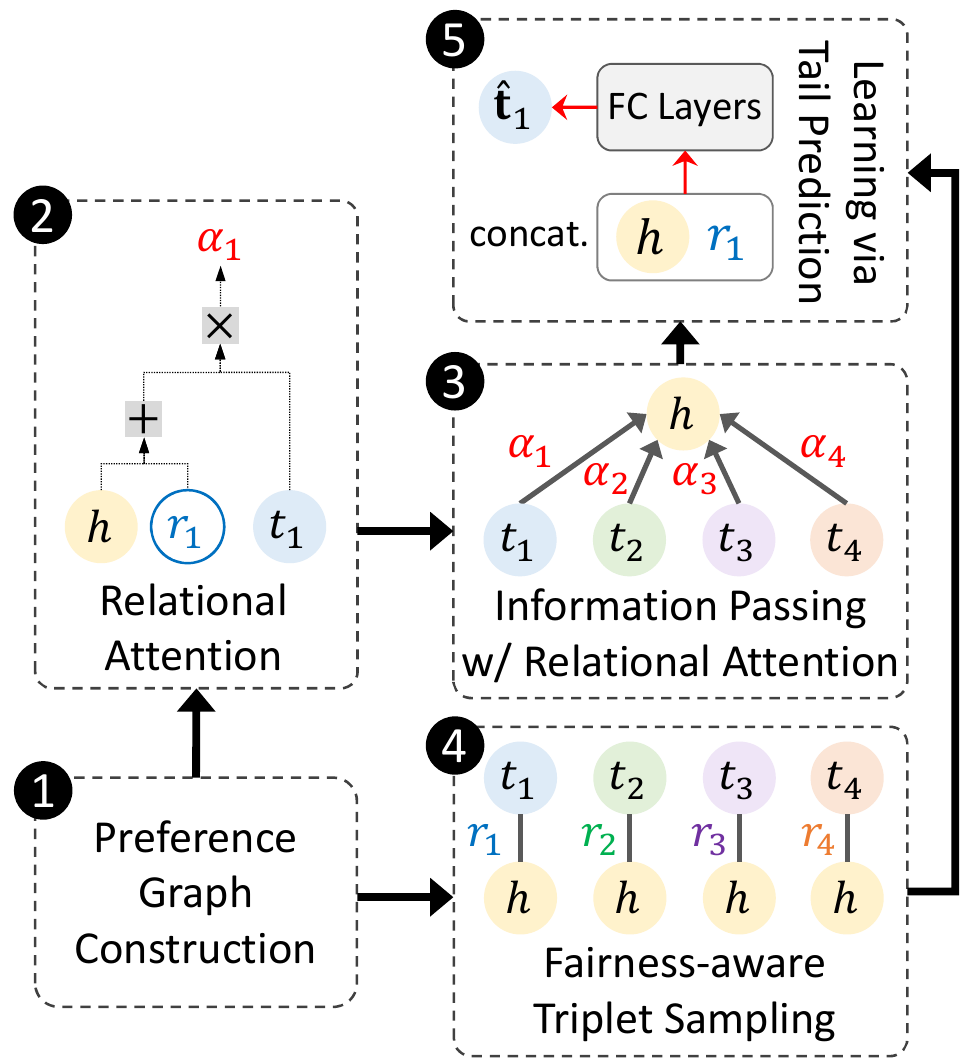}
  \caption{Fairness-aware Preference Graph Embedding.}
  \label{fig:model2}
\end{figure}

The overview of our \texttt{FairSR} framework is presented in Figure~\ref{fig:model}, which consists of two tasks. One is the main task SR, and the other is fairness-aware preference graph embedding (FPGE) shown in Figure~\ref{fig:model2}. When an item subsequence is entered, the cross item-preference learning (CIPL) is used to learn shared features between items and their corresponding entities. Then for the SR part, we design a personalized feature gating to distill effective sequential features, followed by horizontal and vertical convolution mechanisms to extract sequential patterns. By combining sequential features with item-item correlation and global user-item latent factors, SR makes the prediction. For the FPGE part, we first construct a preference graph that depicts the relations between items and attributes. We devise a relational attention-based information passing mechanism to learn entity embeddings. The aim of FPGE is to predict tails based on embeddings of heads and relations. Our framework is trained by alternately optimizing the two tasks.

Here we summarize the novelty of this work into four points. First, we propose a novel task, fairness-aware sequential recommendation, which aims at accurately recommending the next items and simultaneously mitigating the effect of filter bubble in online services. Second, we present a new fairness measure, interaction fairness (IF), to quantify the degree of the filter bubble effect. Our fairness-aware task is to recommend items that can increase IF values. Third, we present a novel multi-task learning-based model, \texttt{FairSR}, to tackle the proposed task. \texttt{FairSR} brings a marriage between personalized sequential feature learning and preference graph embedding in the context of sequential recommendation. Fourth, in the component of preference graph embedding, a new fairness-aware triplet sampling strategy is proposed to ensure that the item embedding learning can encode the fairness across the protected attribute groups.

\subsection{Sequential Feature Learning}
\label{sec-seqfeat}
The basic idea of sequential recommendation is that the recent subsequent items can to some extent influence the adoption of the next items. Hence, existing studies have presented a variety of models to learn the representations of item (sub)sequences, such as convolution neural network~\cite{caser18}, recurrent neural network~\cite{rnnsr18}, self-attention mechanism~\cite{sasrec18}, and feature gating method~\cite{hgn19}. To robustly model the correlation between the current subsequence of items and the next items to be recommended, we propose a joint gating and convolutional subnetwork, which combines \textit{personalized feature gating} and two \textit{convolutional mechanisms}. The personalized feature gating is to select significant sequential features that are positively related to the next item prediction. Convolutional mechanisms aim at modeling sequential patterns from the perspectives of both \textit{union}-level and \textit{point}-level. This section consists of four phases: (a) Item Embedding Layer, (b) Personalized Feature Gating, (c) Convolutional Layers, and (d) Aggregation Layer.

\textbf{User \& Item Embedding Layers.} The input of our model are the one-hot vectors for each user and each item. By feeding one-hot vectors into the user and item embedding layers, where each of which is a hidden layer with dimension $d$, each user and each item can be represented by a low-dimensional real-value dense vector. Let the user embeddings be $\mathbf{U}\in\mathbb{R}^{d\times M}$, and also let the item embeddings be $\mathbf{V}\in\mathbb{R}^{d\times N}$, where $d$ is the embedding dimension and set as $d=32$ by default throughout the paper. Note we denote a user embedding vector and an item embedding vector as $\mathbf{u}\in\mathbb{R}^{d}$ and $\mathbf{v}\in\mathbb{R}^{d}$, respectively. Given the $l$-th item subsequence $\mathcal{S}^u_{1:t}$ of a certain user $u$, its corresponding embeddings can be represented by: 
$$
\mathbf{S}_{u,l}=\begin{bmatrix}
\mid &  & \mid &  & \mid  \\ 
\mathbf{v}_{1} & \cdots & \mathbf{v}_j & \cdots & \mathbf{v}_{t}  \\ 
\mid &  & \mid &  & \mid 
\end{bmatrix}  
$$
where $\mathbf{S}_{u,l}\in\mathbb{R}^{d\times t}$, and $\mathbf{v}_t\in \mathbb{R}^d$ can be retrieved from the item embedding matrix $\mathbf{V}$.

\textbf{Personalized Feature Gating.}
For different users, it is various that which of items in the current sequence are more effective in predicting next ones. The selection of representative items needs to be user-specific. In other words, the feature gating should be personalized to every user. To fulfill feature gating, we take advantage of the \textit{gated linear unit} (GLU)~\cite{glu17}, which is originally devised to detect and propagate effective word embeddings for predicting next word in natural language modeling. We exploit GLU to find what a user $u$ cares about along the sequence of items $\mathbf{S}_{u,l}$. We refer to inner product-based feature gating~\cite{hgn19} to distill item features to users' preferences. The distilled features can be derived through:
\begin{equation}
\mathbf{S}^{F}_{u,l}=\mathbf{S}_{u,l}\otimes \sigma(\mathbf{S}_{u,l}\cdot\mathbf{W}_{g_1} + \mathbf{u} \cdot \mathbf{w}_{g_2}^{\top}),
\end{equation}
where $\mathbf{S}^{F}_{u,l}\in\mathbb{R}^{d\times t}$, $\mathbf{u}\in\mathbb{R}^{d}$ is the embedding of user $u$, $\mathbf{W}_{g_1}\in\mathbb{R}^{t\times t}$, and $\mathbf{w}_{g_2}\in\mathbb{R}^{t}$ are learnable weights.
Each user embedding $\mathbf{u}$ is generated through feeding the one-hot vector of a user into a $d$-dimensional hidden layer, where $d$ is set as $32$ by default.
$\otimes$ is the element-wise product between matrices, and $\sigma$ is the sigmoid function. The distilled sequential feature matrix $\mathbf{S}^{F}_{u,l}$ captures user preferences from past items to the next one, and will be used in follow-up layers.

\textbf{Convolutional Layers.} Equipped with the distilled sequential features, we aim at learning sequential patterns by treating $\mathbf{S}^{F}_{u,l}$ as an image. We exploit convolution filters to search for sequential patterns. Two convolutional filters are adopted. One is \textit{horizontal convolution}, and the other is \textit{vertical convolution}. Horizontal convolution filters ($h\times d$ matrices, $h=2$ by default) are created to find \textit{union}-level sequential patterns, which mean how features of few consecutive items lead to a particular next item. Vertical convolution filters ($t\times 1$ matrices) are devised to learn \textit{point}-level sequential patterns, which refer to how each feature (dimension) distributes over the item subsequence affect the prediction. 

\textit{Horizontal Convolution Filters.} This layer contains $n_h$ horizontal filters $\Psi^j \in \mathbb{R}^{d\times h}$, $1\leq j\leq n_h$, where $h\in \{1,...,t\}$ is the width of a filter. For instance, in Figure~\ref{fig:model}, we can choose $n_h=4$ filters if $t=2$, in which two for each $h$ in $\{1,2\}$. Every filter $\Psi^j$ moves from left to right on $\mathbf{S}^{F}_{u,l}$, and produces convolved values with horizontal dimensions from item $1$ to $t-h+1$ in item subsequence $\mathcal{S}^u_{1:t}$. The resulting vector $\mathbf{c}^{j}$ of horizontal convolution with filter $\Psi_j$ can be obtained by:
\begin{equation}
\mathbf{c}^{j}=[\cdots, \phi_c(\mathbf{S}_{i:i+h-1}\odot\Psi^j), \cdots ],
\end{equation}
where $j=1,2,...,t-h+1$, $\mathbf{S}$ is $\mathbf{S}^{F}_{u,l}$ for simplicity, $\odot$ is the element-wise multiplication, and $\phi_c(\cdot)$ is the activation function for convolutional layers (\textit{tanh} is used by default). Since we have $n_h$ filters, we can derive the resulting matrix $\mathbf{D}\in \mathbb{R}^{(t-h+1)\times n_h}=[\mathbf{c}^{1},\mathbf{c}^{2},\cdots,\mathbf{c}^{n_h}]$.

\textit{Vertical Convolution Filters.} This layer has $n_v$ vertical filters $\Lambda^j \in \mathbb{R}^{1\times t}$, $1\leq j\leq n_v$. Every filter $\Lambda^j$ operates on each embedding dimension of $\mathbf{S}^{F}_{u,l}$, and thus generates $d$ convolved values. Since the derivation of convolved values can be considered as the weighted sum of $\Lambda^j$ on each embedding dimension of $\mathbf{S}^{F}_{u,l}$, we can depict the vertical convolution operation using the inner product, and have the resulting vector $\bar{\mathbf{c}}^j$ by:
\begin{equation}
\bar{\mathbf{c}}^{j}=[\cdots, \sum_{i=1}^{t}\Lambda^j_i\cdot \mathbf{S}_i^{\top}, \cdots ],
\end{equation}
where $\mathbf{S}_i$ is the $i$-th row of matrix $\mathbf{S}^{F}_{u,l}$. In other words, the vertical convolutional layer aggregates the embeddings of past $t$ items through a variety of filters. Since we have $n_v$ filters, we can obtain the resulting matrix $\bar{\mathbf{D}}\in \mathbb{R}^{d\times n_v}=[\bar{\mathbf{c}}^{1},\bar{\mathbf{c}}^{2},\cdots,\bar{\mathbf{c}}^{n_v}]$.

\textbf{Aggregation Layer.} Since the aim is to generate the effective representation for user $u$'s item subsequence $\mathbf{S}^{F}_{u,l}$, we aggregate the convolution results into sequence-level embeddings. We apply \textit{max pooling} to every horizontal convolved vector $\mathbf{c}^j$ and every vertical convolved vector $\bar{\mathbf{c}}^{j}$. By doing so, we can extract the significant features from all values produced by a particular filter. As a result, for $n_h$ horizontal and $n_v$ vertical filters, we can obtain the corresponding aggregated vectors $\mathbf{s}^{hc}_{u,l}\in\mathbb{R}^{n_h}$ and $\mathbf{s}^{vc}_{u,l}\in \mathbb{R}^{n_v}$.
Then such two vectors are concatenated to be the final extracted sequential feature vector, denoted by $\mathbf{s}_{u,l}=[\mathbf{s}^{hc}_{u,l},\mathbf{s}^{vc}_{u,l}]$.
In short, the learned subsequence embedding $\mathbf{s}_{u,l}$ encodes not only user preferences on past items from personalized feature gating, but also captures the sequential patterns in both horizontal and vertical aspects.

\subsection{Cross Item-Preference Learning}
\label{sec-crossl}
Since items can be characterized by how they are interacted by users, we model feature interactions between items and entities in the preference graph. A cross item-preference learning module is developed. We adopt the \textit{cross\&compress unit}~\cite{mkr19} to implement the cross item-preference learning. To be specific, the cross item-preference learning (CIPL) module is to model high-order interactions between items and their corresponding entity features, which captures the ways that users with various attributes express their preferences on items. CIPL can automatically control the cross knowledge transfer between tasks of sequential recommendation and entity embedding learning. Through the learnable weights that bring embeddings of items and entities together in CIPL, the two tasks can affect and complement one another. The CIPL module consists of two main steps, one is \textit{cross}, and the other is \textit{compress}. The \textit{cross} step is devised to produce an interaction matrix between items and entities based on their representations. That said, the interaction matrix models how both sides correlate with each other. The \textit{compress} step utilizes the learned interaction matrix to map the embeddings of items and entities into the same space, i.e., generating the updated embeddings of items and entities at the next layer.

Each item $v\in\mathcal{V}$ is associated with an entity $e$ in the preference graph. We build a $d\times d$ pairwise interaction embedding matrix $\mathbf{C}_l$ for the item embedding $\mathbf{v}_l\in\mathbb{R}^d$ and the entity embedding $\mathbf{e}_l\in\mathbb{R}^d$ at the $l$-th neural network layer. The matrix $\mathbf{C}_l$ depicts the cross feature interactions between every item $v$ and every entity $e$, and can be derived by $\mathbf{C}_l=\mathbf{v}_l\mathbf{e}_l^{\top}\in \mathbb{R}^{d\times d}$, where $d$ is the dimension of hidden layers. Then we generate the next-layer $l+1$ feature vectors of items and entities by projecting the cross feature matrix into their embedding space, given by:
\begin{equation}
\begin{split}
\mathbf{v}_{l+1}&=\mathbf{C}_l\mathbf{w}_l^{\mathcal{V}\mathcal{V}}+\mathbf{C}_l^{\top}\mathbf{w}_l^{\mathcal{E}\mathcal{V}}+\mathbf{b}_l^{\mathcal{V}},\\
\mathbf{e}_{l+1}&=\mathbf{C}_l\mathbf{w}_l^{\mathcal{V}\mathcal{E}}+\mathbf{C}_l^{\top}\mathbf{w}_l^{\mathcal{E}\mathcal{E}}+\mathbf{b}_l^{\mathcal{E}},
\end{split}
\end{equation}
where $\mathbf{w}_l^{\cdot\cdot}\in \mathbb{R}^{d}$ and $\mathbf{b}_l^{\cdot}\in\mathbb{R}^d$ are learnable parameters and bias vectors, respectively. Such a cross learning process can be denoted by: 
\begin{equation}
[\mathbf{v}_{l+1},\mathbf{e}_{l+1}]=\mathcal{C}(\mathbf{v}_l,\mathbf{e}_l).
\end{equation}
We also denote the final updated item embedding matrix as $\bar{V}$. Equipped with cross item-preference learning, the recommendation module can adaptively encode the item/entity knowledge transferred from the preference graph so that the correlation between two tasks can be learned. Note that we consider only lower-level layers in learning cross features. The reason is that lower-layer features can be generally shared by different tasks, and thus are better to be transferred~\cite{ftran1,ftran2}. Higher-layer features tend to be specific to the targeted tasks. The final embeddings of items and entities after $\mathcal{L}$ layers can be represented by: 
\begin{equation}
\begin{split}
\tilde{\mathbf{v}} &= \mathcal{C}^{\mathcal{L}}_v(\mathbf{v},\mathbf{e})\\
\tilde{\mathbf{e}} &= \mathcal{C}^{\mathcal{L}}_e(\mathbf{v},\mathbf{e}),
\end{split}
\end{equation}
respectively, where $\mathcal{L}$ is the number of layers in cross learning and set as $\mathcal{L}=2$ by default.

\subsection{Fairness-aware Preference Graph Embedding Learning}
Preference graph embedding (PGE) aims at encoding both the properties of items and the preferences of users into entity embeddings so that the recommender can be aware of items' and users' knowledge through the abovementioned cross item-preference learning. We also implement the idea of fairness into PGE by considering how items are interacted by users with different attributes when sampling the triplets of the head, relation, and tail in PGE. Below we first present how to construct the preference graph, followed by the learning of entity embeddings with fairness-aware triplet sampling. We will utilize Figure~\ref{fig:model2} to describe the proposed fairness-aware preference graph embedding (FPGE) learning in a step-by-step manner.

\begin{figure*}[!t]
  \centering
  \includegraphics[width=0.85\textwidth]{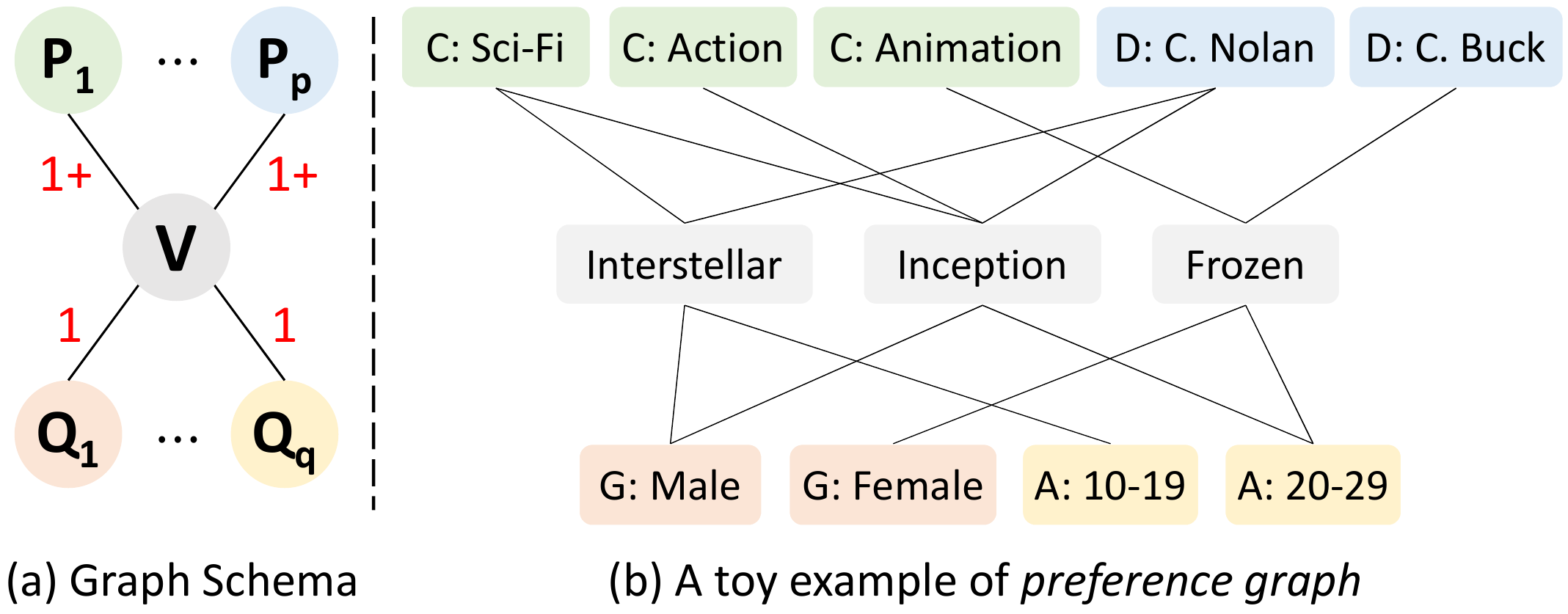}
  \caption{Illustration of preference graphs. (a) The graph schema contains $p$ item properties (P) and $q$ user attributes (Q). In addition, ``\textbf{1+}'' indicates that one item entity can connect to \textit{one or multiple} entites of an item property, and ``\textbf{1}'' means that one item entity can connect to \textit{only one} entity of a user attribute. (b) In the toy example on the movie realm, ``C'', ``D'', ``G'', and ``A'' are categories, director, gender, and age, respectively. The first two are item properties, and the last two are user attributes.}
  \label{fig:pg}
\end{figure*}

\subsubsection{Preference Graph Construction}
Users possessing some attributes may have a higher potential to interact with items with certain properties. We construct a \textit{preference graph} to represent the relationships that user traits interact with item properties. The preference graph is utilized to encode the knowledge on the correlation between user attributes and item properties in the entity embeddings so as to benefit recommendation and provide fairness.

At the first step of FPGE, as shown in Figure~\ref{fig:model2}, we create the preference graph as a tri-partite graph structure, denoted by $\mathcal{G} = (\mathcal{E}, \mathcal{R})$, where $\mathcal{E}$ and $\mathcal{R}$ are the sets of entities and relations, respectively. There are three types of entities in $\mathcal{E}$, including items, user attributes, and item properties. User attributes are those we described in Section~\ref{sec-prob}. Item properties are metadata labels associated on items, such as categories, producers, and origin. There are two kinds of relations in $\mathcal{R}$. One is connecting items to user attributes, and the other is connecting items to item properties. In other words, there are no edges between items, between user attributes, and between item properties. An illustration of the preference graph is shown in Figure~\ref{fig:pg}. The graph schema depicts $p$ item properties (P) and $q$ user attributes (Q). Every item entity can be connected with \textit{only one} entity of a particular user attribute, and \textit{one or multiple} entities of a particular item property. 

In the toy example of Figure~\ref{fig:pg}(b) on the movie realm, which contains $2$ user attributes and $2$ item properties. Each item entity connects to only one gender (G) value and only one age (A) value. Besides, each item entity can connect to one or multiple movie categories and directors. For instance, the movie ``Inception'' is connected to categories ``Sci-Fi'' and ``Action.''

The aim of connecting an item to a user attribute is to not only learn how their correlation, but also enables the incorporation of fairness in PGE learning, which will be described in the following part. However, items are not directly related to user attributes. For each item $v\in\mathcal{V}$ and each attribute $a\in A$, we connect $v$ to only one of $a$'s values $z^a_{\star}\in Z^a$ if $z^a_{\star}$ takes the \textit{major} proportion among users who had ever interacted with $v$. The selection of major attribute value $z^a_{\star}$ for item $v$ can be depicted by: 
\begin{equation}
z^a_{\star} = \arg\max_{z^a\in Z^a} p_v(z^a),
\end{equation}
where $p_v(z^a)$ is the adoption proportion introduced in Section~\ref{sec-prob}.

\subsubsection{PGE Learning}
We consider PGE learning as a kind of knowledge graph embedding~\cite{kgesur17}. In PGE learning, entities in the preference graph can be divided into three types, \textit{head} ($h$), \textit{relation} ($r$), and \textit{tail} ($t$), which can be treated as a triplet $(h,r,t)$. While each item has a corresponding item entity in the preference graph, multiple layers of cross item-preference learning (Section~\ref{sec-crossl}) will first convert the original item embeddings to their latent representations of head entities $\mathbf{h}$, i.e., $\mathbf{h} = \tilde{\mathbf{e}} = \mathcal{C}^{\mathcal{L}}_e(\mathbf{v},\mathbf{e})$. In the meanwhile, the each relation $r$ is also processed through $\mathcal{L}$ fully-connected neural network layers, given by $\mathbf{r}=\mathcal{M}^\mathcal{L}(\mathbf{r}_0)$, where $\mathbf{r}_0$ is the initialized one-hot encoding vector of relation IDs or types, and $\mathcal{M}^\mathcal{L}(\mathbf{x}) = \sigma(\mathbf{W}\mathbf{x} + \mathbf{b})$ is a $\mathcal{L}$-layer fully-connected neural network with trainable weights $\mathbf{W}$, bias $\mathbf{b}$, and the non-linear activation function $\sigma(\cdot)$. 
The process of learning PGE consists of three parts, \textit{information passing}, \textit{relational attention}, and \textit{tail prediction}.

\textbf{Information Passing.} An entity can participate in one or multiple triplets, and thus can be regarded as an intermediate to pass information from one triplet to another in the preference graph. For example, in Figure~\ref{fig:pg}(b), entity ``Frozen'' is involved in two triplets on item properties ``C: Animation'' and ``D: C. Buck'' and two triplets on user attributes ``G: Female'' and ``A: 20-29.'' The last two can be indirectly depicted by the first two through information propagation from the first two to ``Frozen'', which is further passed to the last two. In other words, we encode user attributes and item properties (i.e., neighbors) into item entities in direct and indirect manners. 

Let $\mathcal{N}_h=\{ (h,r,t)|(h,r,t)\in\mathcal{G} \}$ be the set of triplets whose head is $h$, which is the item entity. As shown in Figure~\ref{fig:model2}, at the third step of FPGE, we use the neighbors of entity $h$ to represent $h$ itself via linear combination:
\begin{equation}\label{eq-infop}
\mathbf{h}=\sum_{(h,r,t)\in\mathcal{N}_h} \alpha(h,r,t)\mathbf{t},
\end{equation}
where $\alpha(h,r,t)$ is the \textit{relational attention weight} that determines how the edge $(h,r,t)$ contributes to the representation of head $h$ from tail $t$ conditioned on relation $r$. We can also consider $\alpha(h,r,t)$ as a gate that decides how much information being passed from $t$ to $h$ through $r$.

\textbf{Relational Attention.} We learn the relational attention weights by graph attention mechanisms~\cite{gat18,kgat19} at the second step of FPGE, as shown in Figure~\ref{fig:model2}. The idea is to estimate the distance between entity $h$ and tail $t$ by projecting them into the space of relation $r$, and allow those head-tail pairs with shorter distance to have higher attention weights. Higher Similarity between vectors $\mathbf{h}$ and $\mathbf{t}$ conditioned on relation $r$ leads to much information propagated from $t$ to $h$. The derivation of relational attention is given by:
\begin{equation}
\alpha(h,r,t)=(\mathbf{W}_r\mathbf{t})^{\top}\rho\left( \mathbf{W}_r\mathbf{h}+\mathbf{r} \right),
\end{equation}
where $\rho$ is the non-linear activation function $tanh$, and $\mathbf{W}_r$ is a matrix of learnable weights. Then we further normalize the attention weights over all triplets that entity $h$ participates in based on the softmax function, given by:
\begin{equation}
\alpha(h,r,t)=\frac{\exp(\alpha(h,r,t))}{\sum_{(h,r',t')\in \mathcal{N}_h} \exp(\alpha(h,r',t'))}.
\end{equation}
The resultant attention scores are used in Equation~(\ref{eq-infop}) to highlight which item properties and user attributes provide stronger correlation signals to item entities.

\textbf{Tail Prediction.} At the last step of FPGE, as shown in Figure~\ref{fig:model2}, we concatenate the derived head and relation embedding vectors $\mathbf{h}$ and $\mathbf{r}$ together. By feeding it into a $\mathcal{K}$-layer fully-connected neural network ($\mathcal{K}=1$ by default), our target is to predict the tail embedding $\mathbf{t}$, given by: $\hat{\mathbf{t}} = \mathcal{M}^{\mathcal{K}} (\mathbf{h},\mathbf{r})$, where $\hat{\mathbf{t}}$ is the predicted tail embedding. Since we will eventually put the PGE learning into the final loss function, for each triplet $(h,r,t)$, the normalized inner product is used to generate a score $b$ that measures the goodness of PGE task
, given by:
\begin{equation}
b(h,r,t)=\sigma(\mathbf{t}^{\top}\hat{\mathbf{t}}),
\end{equation}
where $\mathbf{t}$ is the real-value embedding vector of tail $t$.

\subsubsection{Fairness-aware Triplet Sampling}
\label{Sec-fairsamp}
We argue the original random triplet sampling brings \textit{unfair} attribute information in entity embeddings. Past studies pointed out that the bias or unfairness of a recommender comes from how users with particular demographic attributes prefer to interact with some items~\cite{biasatt18,fairhyb18,fbias}, leading to imbalanced distributions of user attributes on items. Hence, random triplet sampling is biased to produce triplets containing entities of popular attributes, which further prohibit recommenders
from being fair. 

To remedy the bias and impose the fairness, we aim at making the item head embeddings be aware of fairness on user attributes through a proposed fairness-based triplet sampling (at the fourth step of FPGE, as shown in Figure~\ref{fig:model2}), which consists of two phases. First, we adjust the probability of each triplet being sampled. Second, we devise an item head-based sampling strategy. Specifically, given the targeted set of attributes $\tilde{A}$ considered for the fairness, for every head $h_v$ of item $v$, we first identify its set of neighboring entities $\mathcal{N}^{\tilde{A}}_{h_v}$, which belong to user attribute $a\in \tilde{A}$. For each tail entity $t_{z^a}\in \mathcal{N}^{\tilde{A}}_{h_v}$, the probability that triplet $(h_v,r,t_{z^a})$ being sampled is proportional to the reciprocal of the number of users with attribute value $z^a$ who had interacted with item $v$ via:
\begin{equation}
\label{eq-samprob}
\pi(h_v,r,t_{z^a})= \frac{1 / |\{ u | \mathcal{A}^{a}(u)=z^a, u\in \mathcal{U}(v) \}|}{\sum_{z^{a'}\in Z^{a}} 1 / |\{ u | \mathcal{A}^{a}(u)=z^{a'}, u\in \mathcal{U}(v) \}|}.
\end{equation}
Take the head entity of item ``Interstellar'' in Figure~\ref{fig:pg}(b) as an example, assume $20$ ``Male'' users $30$ ``10-19'' users had interacted with ``Interstellar'', then we have $\pi(\text{``Interstellar''},r,\text{``Male''})=\frac{1/20}{1/20+1/30}$ and $\pi(\text{``Interstellar''},r,\text{``10-19''})=\frac{1/30}{1/20+1/30}$.

Based on Equation~\ref{eq-samprob}, if item $v$ is frequently interacted by users with attribute value $z^a$, we lower down the probability of the corresponding triplet $(h_v,r,t_{z^a})$ being sampled. By doing so, we can have equal sampling possibility values for entities of all possible values of every attribute. After deriving the probabilities of all triplet $(h_v,r,t_{z^a})$, we normalize them to be in $[0,1]$ for \textit{positive} triplet sampling. The negative triplets are also sampled based on item heads. We randomly select an equal number of non-connected user-attribute tails to be the negative triplets.

\subsection{Modeling Item-Item Correlation}
The next items being recommended is influenced by the correlation between single items in the current item subsequence~\cite{recsurv19}, in addition to their sequential effect. An existing study also extracted rules exhibiting strong correlation between next and current items~\cite{exrule19}. To exploit such item-item correlation, we model the correlation between the updated embeddings (i.e., after cross item-preference learning) of all items $\bar{\mathbf{V}}$ and the original item embeddings in the subsequence $\mathbf{S}_{u,l}$. The inner product is applied to obtain a correlation score, given by:
\begin{equation}
\sum_{\mathbf{v}_j\in \mathbf{S}_{u,l}} \mathbf{v}^{\top}_j\cdot \bar{\mathbf{V}},
\end{equation}
where $\bar{\mathbf{V}}\in\mathbb{R}^{d\times N}$ is the updated item embedding matrix (after cross learning in Section~\ref{sec-crossl}). This score reflects the correlation between each item in $\mathcal{S}^u_{1:t}$ and each candidate item $v\in\mathcal{V}$.

\subsection{Prediction Layer}
The prediction of next items consists of three parts. First, we adopt the conventional matrix factorization~\cite{mfhe16} to capture long-term interests of users. Second, we consider the interaction between users and the sequential features learned in Section~\ref{sec-seqfeat} to model short-term interests of users. Third, the item-item correlation is incorporated in the prediction. For a given $l$-th item subsequence, the predicted score of user $u$ on item $v$ can be represented by:
\begin{equation}
\hat{y}_{u,v}=\bar{\mathbf{u}}^\top\cdot\bar{\mathbf{v}}+\mathbf{s}_{u,l}^\top\cdot\bar{\mathbf{v}}+\sum_{\mathbf{v}_j\in \mathbf{S}_{u,l}} \mathbf{v}^{\top}_j\cdot\bar{\mathbf{v}},
\end{equation}
where $\bar{\mathbf{u}}=\mathcal{M}^{\mathcal{L}}(\mathbf{u})\in \mathbb{R}^d$ is the updated user embedding after fully-connected layers, $\mathbf{u}$ is the input user embedding, $\bar{\mathbf{v}}\in\mathbb{R}^d$ is a column vector of the updated item embedding matrix $\bar{\mathbf{V}}$. We expect the true item $v$ adopted by user $u$ can lead to higher prediction score $\hat{y}_{u,v}$.

\subsection{Model Training}
The overall loss function consists of two parts. One is the loss for user-item prediction in sequential recommendation (SR), and the other is the loss for fairness-aware preference graph embedding (FPGE). We optimize the SR part by Bayesian Personalized Ranking (BPR) objective~\cite{bpr09}, i.e., the pairwise ranking between positive and non-interacted items. In addition, we optimize the FPGE part by increasing the score for all positive triplets while decreasing the score for all negative triplets. The loss function is as follows:
\begin{equation}
\label{eq-loss}
\begin{split}
\mathcal{J}=&\mathcal{J}_{SR}+\mathcal{J}_{FPGE}+\lambda\|\Theta\|^2_2 \\
=&\sum_{(u,v,\mathcal{S}^{u}_l,v')\in\mathcal{D}}-\log\sigma(\hat{y}_{u,v}-\hat{y}_{u,v'}) \\
&-\lambda_1\left(\sum_{(h,r,t)\in\mathcal{G}}b(h,r,t)-\sum_{(h,r',t')\notin\mathcal{G}}b(h,r',t')\right) \\
&+\lambda_2\|\Theta\|^2_2
\end{split}
\end{equation}
where $\mathcal{S}^{u}_{1:t}$ denotes one $t$-length item subsequence of user $u$, $v'$ is one non-interacted item, $\mathcal{D}$ is the entire dataset, $(h,r',t')$ indicates negative triple sampling for efficient training, $\Theta$ contains all learnable parameters in the neural network, and $\lambda_1$ and $\lambda_2$ are the balancing hyperparameters. We utilize Adam~\cite{adam15} to adaptively adjust the learning rate during learning. In each training iteration, since our ultimate goal is for SR, we first repeat train the SR part for $\epsilon$ times ($\epsilon=3$ by default), and train the FPGE part once.

\section{Experiments}
\label{sec-exp}
We conduct a series of extensive experiments to answer the following four evaluation questions. 
\begin{itemize}
\item \textbf{EQ1:} Can the proposed FairSR outperform state-of-the-art models in sequential recommendation?
\item \textbf{EQ2:} Is FairSR able to improve the fairness of recommended items, comparing to other fairness-aware models?
\item \textbf{EQ3:} How does each component of FairSR contribute to the recommendation performance?
\item \textbf{EQ4:} Is FairSR robust to the sensitivity of various hyperparameters?
\end{itemize}

\begin{table*}[!t]
\centering
\caption{Statistics of datasets. Seqs denotes ``sequences.''}
\label{tab:data-stat}
\begin{tabular}{c|c|c|c|c|c}
\hline
 & \# Users & \# Items & \# Interactions & \# PG triplets & \# Seqs \\ \hline
Instagram & 1,007 & 4,687 & 219,690 & 23,784 & 71,216 \\ 
MovieLens-1M & 619 & 2,347 & 125,112 & 14,368 & 83,408 \\ 
Book-Crossing & 384 & 14,910 & 70,696 & 17,232 & 20,232 \\ \hline
\end{tabular}%
\end{table*}

\subsection{Evaluation Setup}
\smallskip\noindent
\textbf{Datasets.} 
Three datasets are employed in the evaluation. (a) \textbf{Instagram}: a user check-in records on locations~\cite{tagv18} at three major urban areas, New York, Los Angeles, and London crawled via Instagram API in 2015. Check-in locations are treated as items. (b) \textbf{MovieLens-1M}\footnote{\url{https://grouplens.org/datasets/movielens/1m/}}: a widely-used benchmark dataset for movie recommendation. (c) \textbf{Book-Crossing}\footnote{\url{http://www2.informatik.uni-freiburg.de/~cziegler/BX/}}: it contains explicit ratings (from $0$ to $10$) of books in the Book-Crossing community. Since MovieLens-1M and Book-Crossing are explicit feedback data, we follow MKR~\cite{mkr19} to convert them into implicit feedback (i.e., $1$ indicates that the user has rated the item and otherwise $0$. The threshold of positive ratings is $4$ for MovieLens-1M $9$ for Book-Crossing. Since the main task is SR and we need user attributes for fairness, we preprocess the datasets by removing users without any attributes and users containing fewer than $4$ interactions with items. The data statistics is summarized in Table~\ref{tab:data-stat}. We have demographic attributes \texttt{gender} and \texttt{age} in both Instagram and Book-Crossing, and \texttt{gender} in MovieLens-1M. We take $10$ years as the range of an attribute value for \texttt{age}. The protected attribute groups are composed of all combinations of attribute values in respective datasets.

\smallskip\noindent
\textbf{Competing Methods.}
We compare several FiarSR with state-of-the-art methods and baselines. Their settings of hyperparameters are tuned by grid search on the validation set.
\begin{itemize}
\item \textbf{Caser}~\footnote{\url{https://github.com/graytowne/caser_pytorch}}~\cite{caser18} is a sequence embedding model that learns sequential features of items through convolution mechanisms. 
\item \textbf{RippleNet}~\footnote{\url{https://github.com/hwwang55/RippleNet}}~\cite{rpnet18} is a memory-network-like approach that
propagates user preferences on items via knowledge graph. 
\item \textbf{SASRec}~\footnote{\url{https://github.com/kang205/SASRec}}~\cite{sasrec18} is a self-attention based sequential model that utilizes the attention mechanism to identify relevant items and their correlation in entire item sequences.
\item \textbf{MKR}~\footnote{\url{https://github.com/hwwang55/MKR}}~\cite{mkr19} (state-of-the-art) a multi-task learning-based model that devises a task interaction learning to combine tasks of user-item matching and knowledge graph embedding.
\item \textbf{HGN}~\footnote{\url{https://github.com/allenjack/HGN}}~\cite{hgn19} (state-of-the-art) a hierarchical gating network that learns the item subsequence embeddings through feature gating in long and short aspects, and models the item-item relations.
\item \textbf{HGN+MKR} (state-of-the-art$+$state-of-the-art): we replace the recommendation module of MKR~\cite{mkr19} with HGN~\cite{hgn19} so that the knowledge graph embedding can be incorporated for sequential recommendation. HGN+MKR is a very strong competitor as being capable of the power of both HGN and MKR.
\item \textbf{FATR}~\footnote{\url{https://github.com/Zziwei/Fairness-Aware_Tensor-Based_Recommendation}}~\cite{ften18} is a fairness-aware tensor-based model that imposes \textit{statistical parity} into item recommendation, i.e., ensuring similar probability distributions of item adoptions for users in different protected attribute groups.
\item \textbf{FairSR-R} replaces fairness-aware triplet sampling with random sampling in FairSR. It serves as a control method to validate whether FairSR leads to fairness.
\item \textbf{FairSR} is the full version of our proposed model.
\end{itemize}

\smallskip\noindent
\textbf{Evaluation Metrics.} 
For SR performance, we adopt in terms of \textit{Precision}$@k$ ($P@k$), \textit{Recall}$@k$ ($R@k$), and \textit{NDCG}$@k$ ($N@k$). To examine whether the recommended items exhibit the proposed \textit{interaction fairness} defined in Section~\ref{sec-prob}, we design the \textit{difference of interaction fairness} (\textit{DIF}@$k$) between the recommended items and the ground truth to be the metric, i.e., $DIF=\hat{IF}-IF$, where $\hat{IF}$ is the IF score generated from recommended items, and $IF$ is the IF score of the corresponding ground truth. Higher positive values of DIF indicate better fairness performance. The negative DIF value of a model means the fairness cannot get improved by that model. 

\begin{table}[!t]
\centering
\caption{Main experimental results of Precision, Recall, and NDCG for sequential recommendation in three datasets. \textbf{Bold} and \underline{underline} indicate the best and the second-best in each metric (column), respectively.}
\label{tab:exp-main}
\begin{tabular}{c|c|c|c|c|c|c|c|c|c}
\hline
 & \multicolumn{3}{c}{Instagram} & \multicolumn{3}{c}{MovieLens-1M} & \multicolumn{3}{c}{Book-Crossing} \\ \hline
 & P@10 & R@10 & N@10 & P@10 & R@10 & N@10 & P@10 & R@10 & N@10 \\ \hline
Caser & 0.0166 & 0.0367 & 0.0199 & 0.0932 & 0.0832 & 0.0953 & 0.0102 & 0.0235 & 0.0125 \\
SASRec & 0.0264 & 0.0414 & 0.0291 & 0.0942 & 0.0862 & 0.0991 & 0.0193 & 0.0491 & 0.0211 \\
RippleNet & 0.0325 & 0.0401 & 0.0389 & 0.1305 & 0.1134 & 0.1387 & 0.0191 & 0.0843 & 0.0274 \\
MKR & 0.0287 & 0.0404 & 0.0281 & 0.1010 & 0.1194 & 0.1165 & 0.0287 & 0.1072 & 0.0294 \\
HGN & 0.0314 & \underline{0.0457} & 0.0321 & 0.1146 & 0.1172 & 0.1317 & 0.0215 & 0.0235 & 0.0319 \\
HGN+MKR & 0.0387 & 0.0402 & 0.0450 & 0.1347 & \underline{0.1292} & \underline{0.1419} & 0.0311 & 0.1131 & 0.0345 \\
FATR & 0.0298 & 0.0326 & 0.0363 & 0.0915 & 0.1045 & 0.0653 & 0.0205 & 0.0954 & 0.0326 \\ \hline
FairSR-R & \textbf{0.0473} & 0.0414 & \textbf{0.0492} & \textbf{0.1462} & \textbf{0.1395} & 0.1414 & \textbf{0.0462} & \underline{0.1383} & \textbf{0.0435} \\ \hline
FairSR & \underline{0.0464} & \textbf{0.0465} & \underline{0.0485} & \underline{0.1389} & 0.1271 & \textbf{0.1449} & \underline{0.0408} & \textbf{0.1452} & \underline{0.0416} \\ \hline
\end{tabular}%
\end{table}

\smallskip\noindent
\textbf{Experimental Settings.}
The ratio of training, validation, and test sets is $6:2:2$. We repeat every experiment $10$ times, and report the average results. We follow existing studies~\cite{caser18,hgn19} to fix the subsequence length $t=5$ and $L=8$, i.e., future length $g=L-t=3$, by default, and will report the results of varying $t$ and $g$. In FairSR, we apply a grid search for selecting proper hyperparameters using the validation set. Eventually we set $\lambda_1=1$, $\lambda_2=10^{-6}$, and $\epsilon=3$ by default for all datasets. 
We examine how different hyperparameters affect the performance in the following. All experiments are conducted with PyTorch running on GPU machines (Nvidia GeForce GTX 1080 Ti).

\subsection{Experimental Results}
\label{sec-expres}
\smallskip\noindent
\textbf{SR Performance Comparison.}
To answer \textbf{EQ1}, we present the results on SR performance shown in Table~\ref{tab:exp-main}. We have the following findings. First, FairSR and FairSR-R consistently outperform the state-of-the-art methods and baselines in terms of precision, recall, and NDCG. The superiority indicates that although FairSR is originally devised to bring fairness into SR, it can still maintain and even improve the SR performance. Second, FairSR-R is slightly better than FairSR. Such a result is expectable because FairSR sacrifices popular (i.e., biased) attributes and items to bring fairness in recommended items. Third, among state-of-the-art methods, HGN+MKR is the most competitive, but is still worse than FairSR. This result implies that although HGN+MKR can incorporate knowledge graphs among items into SR, the modeling of convolution mechanisms, relational attention, personalized feature gating, and fairness-aware sampling in FairSR can further improve the performance. Fourth, among the metrics, FairSR-R generates more significant improvement on Precision, 22.2\%, 8.5\%, and 48.6\% improvement over the most competitive state-of-the-art HGN+MKR. Such results prove the top-$k$ items recommended by FairSR-R can accurately capture user preferences.

\begin{table}[!t]
\centering
\caption{Main experimental results of the difference of interaction fairness (DIF) in three datasets. \textbf{Bold} and \underline{underline} indicate the best and the second-best in each metric (column), respectively.}
\label{tab:exp-fair}
\begin{tabular}{c|c|c|c}
\hline
 & \multicolumn{3}{c}{DIF@10} \\ \hline
 & Instagram & MovieLens-1M & Book-Crossing \\ \hline
Caser & 0.2293 & 0.0103 & 0.1031 \\
SASRec & 0.1835 & \underline{0.0158} & 0.1246 \\
RippleNet & 0.1331 & -0.0039 & 0.0526 \\
MKR & 0.1390 & -0.0759 & 0.0684 \\
HGN & 0.3074 & 0.0098 & 0.1138 \\
HGN+MKR & 0.4583 & -0.0823 & 0.0710 \\
FATR & \underline{0.6632} & 0.0042 & \underline{0.1943} \\ \hline
FairSR-R & 0.6015 & -0.0964 & 0.1572 \\ \hline
FairSR & \textbf{0.6979} & \textbf{0.0395} & \textbf{0.2217} \\ \hline
\end{tabular}%
\end{table}

\begin{figure*}[!t]
  \centering
  \includegraphics[width=1.0\textwidth]{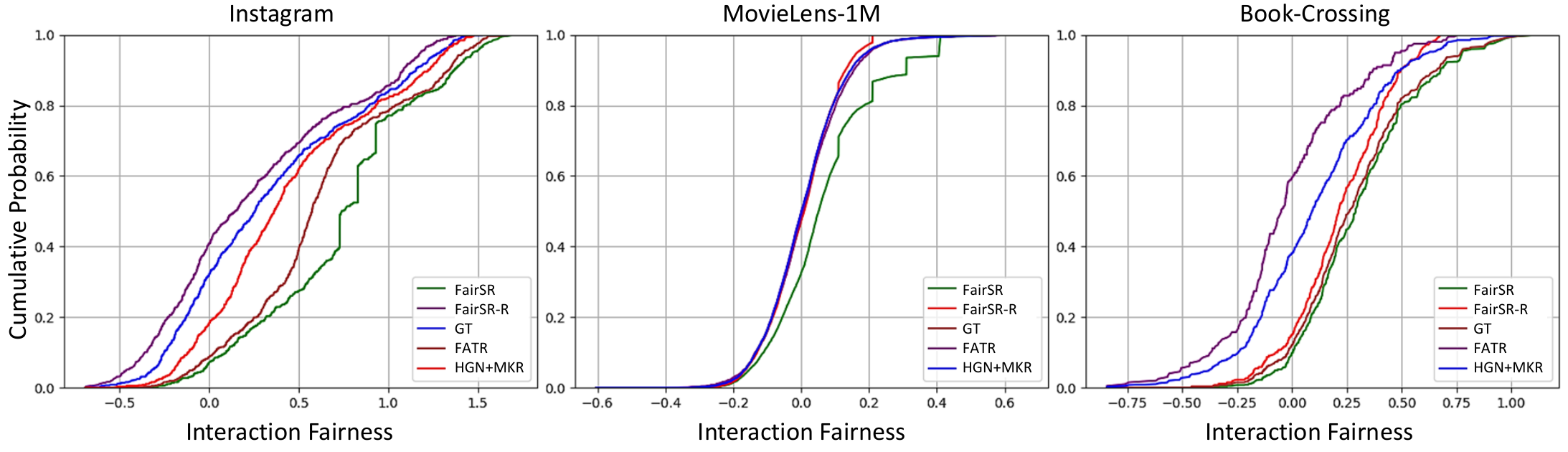}
  \caption{Results on distributions of cumulative probability for IF, i.e., accumulated over users's IF values, in Instagram data. 
  }
  \label{fig:cdf}
\end{figure*}

\smallskip\noindent
\textbf{Fairness Comparison.}
To answer \textbf{EQ2}, we report the results on interaction fairness based on different models, as shown in Table~\ref{tab:exp-fair}. We can find that FairSR consistently produces the highest DIF scores over FairSR-R, the fairness baseline FATR, and other SR competing methods, and the superiority is obviously significant. In terms of DIF scores, the most competitive method is FATR. Although FATR leads to the second-best in Instagram and Book-Crossing datasets, our FairSR can still generate the highest DIF scores in all datasets, and FATR apparently has unsatisfying SR performance. While DIF is an aggregated statistic over users, we further plot the cumulative probability (y-axis) distributions on IF values (x-axis), i.e., accumulated by users' IF values from low to high. We compare IF cumulative probability distributions of FairSR, FairSR-R, the ground truth (GT), and HGN+MKR, and the results are shown in Figure~\ref{fig:cdf}.
All of these results indicate that FairSR is able to not only improve SR performance, but also largely mitigate the unfairness coming from biased interactions between items and user attributes. In addition, the results also imply the proposed fairness-aware preference graph embedding truly takes effect. Moreover, it also provides strong evidence and positive feasibility on having satisfying and fair SR. On the other hand, the DIF values of competing methods are relatively low. We think the reason is existing SR tends to learn user preferences on popular items, which leads to biased user-item interactions. Although the fairness-aware RS method FATR can to some extent improve the DIF values, it still falls behind our proposed FairSR.


\smallskip\noindent
\textbf{Ablation Analysis.}
To answer \textbf{EQ3}, we examine the contribution of each component in FairSR. In this experiment, we compare the FairSR full model with those replacing fairness-aware sampling with random sampling (\textbf{-FS}, i.e., FairSR-R), removing rational attention (\textbf{-RA}), removing convolution mechanisms (\textbf{-Conv}), removing personalized feature gating (\textbf{-PFG}), and removing FPGE (\textbf{-FPGE}). We also remove both personalized feature gating and FPGE (\textbf{-PFG\&FPGE}) from the full model since they are two main designs in this work. The results in Table~\ref{tab:exp-ablation} show that each component truly contributes to the full model. FPGE contributes most, i.e., leading to the largest performance loss, indicating that FairSR highly relies on FPGE to encode how users with different attributes interact with items. In addition, the model is significantly damaged if both PFG and FPGE are removed (\textbf{-PFG\&FPGE}). Such a result again verifies the usefulness of our main technical designs. By looking into the details on the removal of each component in Table~\ref{tab:exp-ablation}, we can have the following two deeper insights. First, removing the fairness-aware sampling (\textbf{-FS}) from the full FairSR model can improve the performance of sequential recommendation in the cases of NDCG@10 on Instagram and NDCG@10 on Book-Crossing. We think such performance drop of ``\textbf{-FS}'' is reasonable and acceptable. FairSR aims at striking a balance between SR performance and interaction fairness. The model with fairness-aware sampling makes the training totally focus on generating good SR performance, which is reflected on the results of ``\textbf{-FS}''. Second, we can find that in the case of Recall@10 on MovieLens, the SR performance gets slightly improved the FairSR model without relational attention (\textbf{-RA}). The potential reason can be that there is only one attribute ``gender'' considered in the construction of the preference graph. Hence, the relational attention mechanism cannot work well to differentiate items associated with various properties in the preference graph.

\begin{table*}[!t]
\centering
\caption{Results on Recall and NDCG for ablation analysis.}
\label{tab:exp-ablation}
\begin{tabular}{l|c|c|c|c|c|c}
\hline
 & \multicolumn{2}{c|}{Instagram} & \multicolumn{2}{c|}{MovieLens-1M} & \multicolumn{2}{c}{Book-Crossing} \\ \hline
 & R@10 & N@10 & R@10 & N@10 & R@10 & N@10 \\ \hline
FairSR & \textbf{0.0465} & 0.0485 & 0.1271 & \textbf{0.1449} & \textbf{0.1249} & 0.0416 \\ 
-FS & 0.0414 & 0.0492 & 0.1156 & 0.1414 & 0.1114 & \textbf{0.0435} \\ 
-RA & 0.0394 & 0.0469 & \textbf{0.1295} & 0.1379 & 0.1172 & 0.0381 \\ 
-Conv & 0.0439 & 0.0479 & 0.1093 & 0.1389 & 0.0962 & 0.0401 \\ 
-PFG & 0.0414 & \textbf{0.0516} & 0.1194 & 0.1487 & 0.0917 & 0.0372 \\ 
-FPGE & 0.0375 & 0.0294 & 0.1039 & 0.1075 & 0.0472 & 0.0199 \\ 
-PFG\&FPGE & 0.0332 & 0.0340 & 0.0914 & 0.1064 & 0.0351 & 0.0128 \\ \hline
\end{tabular}%
\end{table*}



\begin{figure}[!t]
  \centering
  \includegraphics[width=\columnwidth]{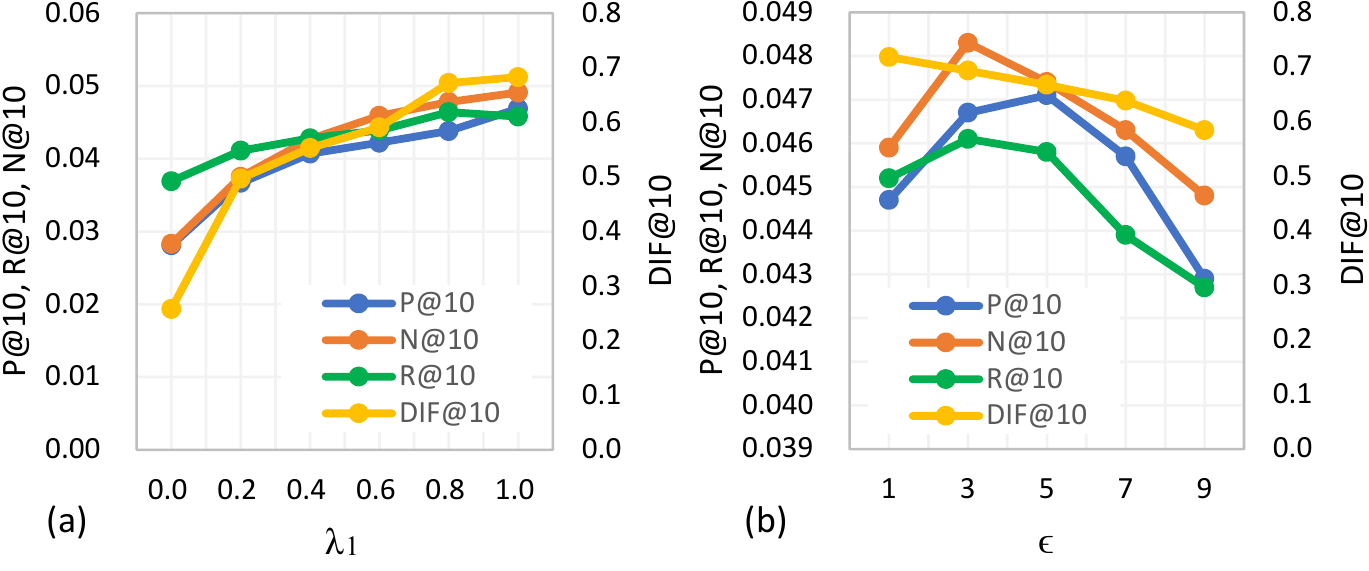}
  \caption{The studies of hyperparameter sensivity of FairSR for (a) the weight of FPGE loss $\lambda_1$ and (b) the SR training frequency $\epsilon$.}
  \label{fig:hsen}
\end{figure}

\smallskip\noindent
\textbf{Hyperparameter Sensitivity.}
To answer \textbf{EQ4}, we present the effect of three hyperparameters: the contribution of FPGE loss $\lambda_1$ in Equation~\ref{eq-loss}, the frequency of SR training $\epsilon$ in each iteration, and the lengths of training and testing sequences $t$ and $g$. 
The results are reported in Figure~\ref{fig:hsen} and Table~\ref{tab:exp-length}. We can first find that higher $\lambda_1$ values lead to better SR performance and higher interaction fairness. Such a result implies the proposed fairness-aware preference graph embedding is able to bring a positive effect on both SR and fairness. As for $\epsilon$, paying too much attention to the SR task (i.e., larger $\epsilon$) weakens the contribution from the FPGE task that simultaneously models interactions between items and attributes and alleviates unfairness via triplet sampling. A proper $\epsilon$ is suggested to be $3$ that strikes a balance between SR and FPGE in SR performance and fairness. On the other hand, we can have the following observations. The length with $t=5$ \& $g=3$ produces better performance. Both increasing $t$ with fixed $g$ and increasing $g$ with fixed $t$ lead to performance improvement. This indicates that we need more training items (higher $t$) to learn short-term user interest, and the given item sequence can determine multiple preferred items (higher $g$).

\begin{table*}[!t]
\centering
\caption{The effect of the sequence length $t$ and $g$ for FairSR.}
\label{tab:exp-length}
\begin{tabular}{c|c|c|c|c|c|c}
\hline
 & \multicolumn{2}{c|}{Instagram} & \multicolumn{2}{c|}{MovieLens-1M} & \multicolumn{2}{c}{Book-Crossing} \\ \hline
 & R@10 & N@10 & R@10 & N@10 & R@10 & N@10 \\ \hline
$t$=3 \& $g$=1 & 0.0375 & 0.0342 & 0.1194 & 0.1328 & 0.1134 & 0.0383 \\ 
$t$=3 \& $g$=2 & 0.0386 & 0.0353 & 0.1202 & 0.1377 & 0.1159 & 0.0394 \\ 
$t$=3 \& $g$=3 & 0.0409 & 0.0373 & 0.1214 & 0.1401 & 0.1217 & 0.0417 \\ 
$t$=5 \& $g$=1 & 0.0403 & 0.0385 & 0.1235 & 0.1392 & 0.1207 & 0.0395 \\ 
$t$=5 \& $g$=2 & 0.0451 & 0.0436 & 0.1229 & 0.1413 & 0.1221 & 0.0403 \\ 
$t$=5 \& $g$=3 & \textbf{0.0489} & \textbf{0.0473} & 0.1247 & \textbf{0.1459} & \textbf{0.1253} & \textbf{0.0424} \\ 
$t$=8 \& $g$=1 & 0.0433 & 0.0423 & 0.1301 & 0.1411 & 0.1229 & 0.0403 \\ 
$t$=8 \& $g$=2 & 0.0468 & 0.0439 & 0.1319 & 0.1427 & 0.1217 & 0.0413 \\ 
$t$=8 \& $g$=3 & 0.0479 & 0.0467 & \textbf{0.1337} & 0.1449 & 0.1244 & 0.0417 \\ \hline
\end{tabular}%
\end{table*}

\section{Conclusion}
\label{sec-conclude}
While the filter bubble effect widely happens to online services, it is crucial to have it considered in recommender systems. To alleviate the filter bubble effect, this paper proposes and solves a novel fairness-aware sequential recommendation task. We define a new metric, interaction fairness, which reflects the degree that items are equally interacted by users with different protected attribute groups. A multi-task learning approach, FairSR, is developed to not only learn personalized sequential features, but also model fairness through embedding items and attributes into the same space via a fair triplet sampling. Experimental results on three datasets prove the effectiveness of FairSR over state-of-the-art SR models and other fairness-aware RS, and exhibit promising interaction fairness. 

There are several future extensions based on FairSR. First, the current preference graph presumes that there is no correlation between item properties and between user attributes. We are seeking to automatically learn their underlying relationships. Second, scalability is an essential issue for recommender systems. The next step is to have FairSR scalable to millions of interactions between users and items. Third, in a realistic setting of recommender systems, new users and new items are continuously coming. Hence, we plan to have FairSR to be capable of inductive learning. 

\begin{acks}
This work is supported by Ministry of Science and Technology (MOST) of Taiwan under grants 110-2221-E-006-136-MY3, 110-2221-E-006-001, and and 110-2634-F-002-051.
\end{acks}

\bibliographystyle{ACM-Reference-Format}
\bibliography{9-References}


\begin{thebibliography}{49}


\ifx \showCODEN    \undefined \def \showCODEN     #1{\unskip}     \fi
\ifx \showDOI      \undefined \def \showDOI       #1{#1}\fi
\ifx \showISBNx    \undefined \def \showISBNx     #1{\unskip}     \fi
\ifx \showISBNxiii \undefined \def \showISBNxiii  #1{\unskip}     \fi
\ifx \showISSN     \undefined \def \showISSN      #1{\unskip}     \fi
\ifx \showLCCN     \undefined \def \showLCCN      #1{\unskip}     \fi
\ifx \shownote     \undefined \def \shownote      #1{#1}          \fi
\ifx \showarticletitle \undefined \def \showarticletitle #1{#1}   \fi
\ifx \showURL      \undefined \def \showURL       {\relax}        \fi
\providecommand\bibfield[2]{#2}
\providecommand\bibinfo[2]{#2}
\providecommand\natexlab[1]{#1}
\providecommand\showeprint[2][]{arXiv:#2}

\bibitem[\protect\citeauthoryear{Abdollahpouri}{Abdollahpouri}{2019}]%
        {dpbias19}
\bibfield{author}{\bibinfo{person}{Himan Abdollahpouri}.}
  \bibinfo{year}{2019}\natexlab{}.
\newblock \showarticletitle{Popularity Bias in Ranking and Recommendation}. In
  \bibinfo{booktitle}{\emph{Proceedings of the 2019 AAAI/ACM Conference on AI,
  Ethics, and Society}} \emph{(\bibinfo{series}{AIES '19})}.
  \bibinfo{pages}{529--530}.
\newblock


\bibitem[\protect\citeauthoryear{Abdollahpouri, Mansoury, Burke, and
  Mobasher}{Abdollahpouri et~al\mbox{.}}{2019}]%
        {fbias}
\bibfield{author}{\bibinfo{person}{Himan Abdollahpouri},
  \bibinfo{person}{Masoud Mansoury}, \bibinfo{person}{Robin Burke}, {and}
  \bibinfo{person}{Bamshad Mobasher}.} \bibinfo{year}{2019}\natexlab{}.
\newblock \bibinfo{title}{The Impact of Popularity Bias on Fairness and
  Calibration in Recommendation}.
\newblock
\newblock
\showeprint[arxiv]{1910.05755}~[cs.IR]


\bibitem[\protect\citeauthoryear{Beutel, Chen, Doshi, Qian, Wei, Wu, Heldt,
  Zhao, Hong, Chi, and et~al.}{Beutel et~al\mbox{.}}{2019}]%
        {fpwc19}
\bibfield{author}{\bibinfo{person}{Alex Beutel}, \bibinfo{person}{Jilin Chen},
  \bibinfo{person}{Tulsee Doshi}, \bibinfo{person}{Hai Qian},
  \bibinfo{person}{Li Wei}, \bibinfo{person}{Yi Wu}, \bibinfo{person}{Lukasz
  Heldt}, \bibinfo{person}{Zhe Zhao}, \bibinfo{person}{Lichan Hong},
  \bibinfo{person}{Ed~H. Chi}, {and} \bibinfo{person}{et al.}}
  \bibinfo{year}{2019}\natexlab{}.
\newblock \showarticletitle{Fairness in Recommendation Ranking through Pairwise
  Comparisons}. In \bibinfo{booktitle}{\emph{Proceedings of the 25th ACM SIGKDD
  International Conference on Knowledge Discovery \& Data Mining}}
  \emph{(\bibinfo{series}{KDD '19})}. \bibinfo{pages}{2212--2220}.
\newblock


\bibitem[\protect\citeauthoryear{Bose and Hamilton}{Bose and Hamilton}{2019}]%
        {cfgrl19}
\bibfield{author}{\bibinfo{person}{Avishek Bose} {and} \bibinfo{person}{William
  Hamilton}.} \bibinfo{year}{2019}\natexlab{}.
\newblock \showarticletitle{Compositional Fairness Constraints for Graph
  Embeddings}. In \bibinfo{booktitle}{\emph{Proceedings of the 36th
  International Conference on Machine Learning}} \emph{(\bibinfo{series}{ICML
  '19}, Vol.~\bibinfo{volume}{97})}. \bibinfo{pages}{715--724}.
\newblock


\bibitem[\protect\citeauthoryear{Cai, Wang, Wu, Wang, and San.}{Cai
  et~al\mbox{.}}{2021}]%
        {catesr21}
\bibfield{author}{\bibinfo{person}{Renqin Cai}, \bibinfo{person}{Hongning
  Wang}, \bibinfo{person}{Jibang Wu}, \bibinfo{person}{Chong Wang}, {and}
  \bibinfo{person}{Aidan San.}} \bibinfo{year}{2021}\natexlab{}.
\newblock \showarticletitle{Category-aware Collaborative Sequential
  Recommendation}. In \bibinfo{booktitle}{\emph{Proceedings of the 44th
  International ACM SIGIR Conference on Research and Development in Information
  Retrieval}} \emph{(\bibinfo{series}{SIGIR '21})}.
\newblock


\bibitem[\protect\citeauthoryear{Cao, Wang, He, Hu, and Chua}{Cao
  et~al\mbox{.}}{2019}]%
        {jkgr19}
\bibfield{author}{\bibinfo{person}{Yixin Cao}, \bibinfo{person}{Xiang Wang},
  \bibinfo{person}{Xiangnan He}, \bibinfo{person}{Zikun Hu}, {and}
  \bibinfo{person}{Tat-Seng Chua}.} \bibinfo{year}{2019}\natexlab{}.
\newblock \showarticletitle{Unifying Knowledge Graph Learning and
  Recommendation: Towards a Better Understanding of User Preferences}. In
  \bibinfo{booktitle}{\emph{The World Wide Web Conference}}
  \emph{(\bibinfo{series}{WWW '19})}. \bibinfo{pages}{151--161}.
\newblock


\bibitem[\protect\citeauthoryear{Chen, Zhang, Ma, Liu, and Ma}{Chen
  et~al\mbox{.}}{2020}]%
        {jnskr20}
\bibfield{author}{\bibinfo{person}{Chong Chen}, \bibinfo{person}{Min Zhang},
  \bibinfo{person}{Weizhi Ma}, \bibinfo{person}{Yiqun Liu}, {and}
  \bibinfo{person}{Shaoping Ma}.} \bibinfo{year}{2020}\natexlab{}.
\newblock \showarticletitle{Jointly Non-Sampling Learning for Knowledge Graph
  Enhanced Recommendation}. In \bibinfo{booktitle}{\emph{Proceedings of the
  43rd International ACM SIGIR Conference on Research and Development in
  Information Retrieval}} \emph{(\bibinfo{series}{SIGIR '20})}.
  \bibinfo{pages}{189--198}.
\newblock


\bibitem[\protect\citeauthoryear{Dauphin, Fan, Auli, and Grangier}{Dauphin
  et~al\mbox{.}}{2017}]%
        {glu17}
\bibfield{author}{\bibinfo{person}{Yann~N. Dauphin}, \bibinfo{person}{Angela
  Fan}, \bibinfo{person}{Michael Auli}, {and} \bibinfo{person}{David
  Grangier}.} \bibinfo{year}{2017}\natexlab{}.
\newblock \showarticletitle{Language Modeling with Gated Convolutional
  Networks}. In \bibinfo{booktitle}{\emph{Proceedings of the 34th International
  Conference on Machine Learning - Volume 70}} \emph{(\bibinfo{series}{ICML
  '17})}. \bibinfo{pages}{933--941}.
\newblock


\bibitem[\protect\citeauthoryear{Ekstrand, Tian, Azpiazu, Ekstrand, Anuyah,
  McNeill, and Pera}{Ekstrand et~al\mbox{.}}{2018}]%
        {biasatt18}
\bibfield{author}{\bibinfo{person}{Michael~D. Ekstrand}, \bibinfo{person}{Mucun
  Tian}, \bibinfo{person}{Ion~Madrazo Azpiazu}, \bibinfo{person}{Jennifer~D.
  Ekstrand}, \bibinfo{person}{Oghenemaro Anuyah}, \bibinfo{person}{David
  McNeill}, {and} \bibinfo{person}{Maria~Soledad Pera}.}
  \bibinfo{year}{2018}\natexlab{}.
\newblock \showarticletitle{All The Cool Kids, How Do They Fit In?: Popularity
  and Demographic Biases in Recommender Evaluation and Effectiveness}. In
  \bibinfo{booktitle}{\emph{Proceedings of the 1st Conference on Fairness,
  Accountability and Transparency}}, Vol.~\bibinfo{volume}{81}.
  \bibinfo{pages}{172--186}.
\newblock


\bibitem[\protect\citeauthoryear{Farnadi, Kouki, Thompson, Srinivasan, and
  Getoor}{Farnadi et~al\mbox{.}}{2018}]%
        {fairhyb18}
\bibfield{author}{\bibinfo{person}{Golnoosh Farnadi}, \bibinfo{person}{Pigi
  Kouki}, \bibinfo{person}{Spencer~K. Thompson}, \bibinfo{person}{Sriram
  Srinivasan}, {and} \bibinfo{person}{Lise Getoor}.}
  \bibinfo{year}{2018}\natexlab{}.
\newblock \showarticletitle{A Fairness-aware Hybrid Recommender System}. In
  \bibinfo{booktitle}{\emph{Proceedings of 2nd FATREC Workshop: Responsible
  Recommendation}}.
\newblock


\bibitem[\protect\citeauthoryear{Feng, Hsu, Li, Yeh, and Lin}{Feng
  et~al\mbox{.}}{2019}]%
        {marine19}
\bibfield{author}{\bibinfo{person}{Ming-Han Feng}, \bibinfo{person}{Chin-Chi
  Hsu}, \bibinfo{person}{Cheng-Te Li}, \bibinfo{person}{Mi-Yen Yeh}, {and}
  \bibinfo{person}{Shou-De Lin}.} \bibinfo{year}{2019}\natexlab{}.
\newblock \showarticletitle{MARINE: Multi-Relational Network Embeddings with
  Relational Proximity and Node Attributes}. In \bibinfo{booktitle}{\emph{The
  World Wide Web Conference}} \emph{(\bibinfo{series}{WWW '19})}.
  \bibinfo{pages}{470--479}.
\newblock


\bibitem[\protect\citeauthoryear{Geyik, Ambler, and Kenthapadi}{Geyik
  et~al\mbox{.}}{2019}]%
        {frank19}
\bibfield{author}{\bibinfo{person}{Sahin~Cem Geyik}, \bibinfo{person}{Stuart
  Ambler}, {and} \bibinfo{person}{Krishnaram Kenthapadi}.}
  \bibinfo{year}{2019}\natexlab{}.
\newblock \showarticletitle{Fairness-Aware Ranking in Search and Recommendation
  Systems with Application to LinkedIn Talent Search}. In
  \bibinfo{booktitle}{\emph{Proceedings of the 25th ACM SIGKDD International
  Conference on Knowledge Discovery \& Data Mining}}
  \emph{(\bibinfo{series}{KDD '19})}. \bibinfo{pages}{2221--2231}.
\newblock


\bibitem[\protect\citeauthoryear{He, Zhang, Kan, and Chua}{He
  et~al\mbox{.}}{2016}]%
        {mfhe16}
\bibfield{author}{\bibinfo{person}{Xiangnan He}, \bibinfo{person}{Hanwang
  Zhang}, \bibinfo{person}{Min-Yen Kan}, {and} \bibinfo{person}{Tat-Seng
  Chua}.} \bibinfo{year}{2016}\natexlab{}.
\newblock \showarticletitle{Fast Matrix Factorization for Online Recommendation
  with Implicit Feedback}. In \bibinfo{booktitle}{\emph{Proceedings of the 39th
  International ACM SIGIR Conference on Research and Development in Information
  Retrieval}} \emph{(\bibinfo{series}{SIGIR '16})}. \bibinfo{pages}{549--558}.
\newblock


\bibitem[\protect\citeauthoryear{Hidasi and Karatzoglou}{Hidasi and
  Karatzoglou}{2018}]%
        {rnnsr18}
\bibfield{author}{\bibinfo{person}{Bal\'{a}zs Hidasi} {and}
  \bibinfo{person}{Alexandros Karatzoglou}.} \bibinfo{year}{2018}\natexlab{}.
\newblock \showarticletitle{Recurrent Neural Networks with Top-k Gains for
  Session-Based Recommendations}. In \bibinfo{booktitle}{\emph{Proceedings of
  the 27th ACM International Conference on Information and Knowledge
  Management}} \emph{(\bibinfo{series}{CIKM '18})}. \bibinfo{pages}{843--852}.
\newblock


\bibitem[\protect\citeauthoryear{Hidasi, Karatzoglou, Baltrunas, and
  Tikk}{Hidasi et~al\mbox{.}}{2016}]%
        {gru4rec16}
\bibfield{author}{\bibinfo{person}{Balazs Hidasi}, \bibinfo{person}{Alexandros
  Karatzoglou}, \bibinfo{person}{Linas Baltrunas}, {and}
  \bibinfo{person}{Domonkos Tikk}.} \bibinfo{year}{2016}\natexlab{}.
\newblock \showarticletitle{Session-based Recommendations with Recurrent Neural
  Networks}. In \bibinfo{booktitle}{\emph{Proceedings of International
  Conference on Learning Representations}} \emph{(\bibinfo{series}{ICLR '16})}.
\newblock


\bibitem[\protect\citeauthoryear{Hsu and Li}{Hsu and Li}{2021}]%
        {retagnn21}
\bibfield{author}{\bibinfo{person}{Cheng Hsu} {and} \bibinfo{person}{Cheng-Te
  Li}.} \bibinfo{year}{2021}\natexlab{}.
\newblock \showarticletitle{RetaGNN: Relational Temporal Attentive Graph Neural
  Networks for Holistic Sequential Recommendation}. In
  \bibinfo{booktitle}{\emph{Proceedings of the Web Conference}}
  \emph{(\bibinfo{series}{WWW '21})}.
\newblock


\bibitem[\protect\citeauthoryear{Jiang, Li, Chen, and Wang}{Jiang
  et~al\mbox{.}}{2018}]%
        {accu18}
\bibfield{author}{\bibinfo{person}{Jyun-Yu Jiang}, \bibinfo{person}{Cheng-Te
  Li}, \bibinfo{person}{Yian Chen}, {and} \bibinfo{person}{Wei Wang}.}
  \bibinfo{year}{2018}\natexlab{}.
\newblock \showarticletitle{Identifying Users behind Shared Accounts in Online
  Streaming Services}. In \bibinfo{booktitle}{\emph{The 41st International ACM
  SIGIR Conference on Research \& Development in Information Retrieval}}
  \emph{(\bibinfo{series}{SIGIR '18})}. \bibinfo{pages}{65--74}.
\newblock


\bibitem[\protect\citeauthoryear{{Kang} and {McAuley}}{{Kang} and
  {McAuley}}{2018}]%
        {sasrec18}
\bibfield{author}{\bibinfo{person}{W. {Kang}} {and} \bibinfo{person}{J.
  {McAuley}}.} \bibinfo{year}{2018}\natexlab{}.
\newblock \showarticletitle{Self-Attentive Sequential Recommendation}. In
  \bibinfo{booktitle}{\emph{2018 IEEE International Conference on Data Mining
  (ICDM)}}. \bibinfo{pages}{197--206}.
\newblock


\bibitem[\protect\citeauthoryear{Kingma and Ba}{Kingma and Ba}{2015}]%
        {adam15}
\bibfield{author}{\bibinfo{person}{Diederik~P. Kingma} {and}
  \bibinfo{person}{Jimmy~Lei Ba}.} \bibinfo{year}{2015}\natexlab{}.
\newblock \showarticletitle{Adam: A Method for Stochastic Optimization}. In
  \bibinfo{booktitle}{\emph{Proceedings of International Conference on Learning
  Representations}} \emph{(\bibinfo{series}{ICLR '15})}.
\newblock


\bibitem[\protect\citeauthoryear{Kumar, Zhang, and Leskovec}{Kumar
  et~al\mbox{.}}{2019}]%
        {jodie19}
\bibfield{author}{\bibinfo{person}{Srijan Kumar}, \bibinfo{person}{Xikun
  Zhang}, {and} \bibinfo{person}{Jure Leskovec}.}
  \bibinfo{year}{2019}\natexlab{}.
\newblock \showarticletitle{Predicting Dynamic Embedding Trajectory in Temporal
  Interaction Networks}. In \bibinfo{booktitle}{\emph{Proceedings of the 25th
  ACM SIGKDD International Conference on Knowledge Discovery \& Data Mining}}
  \emph{(\bibinfo{series}{KDD '19})}. \bibinfo{pages}{1269--1278}.
\newblock


\bibitem[\protect\citeauthoryear{Long, Cao, Wang, and Jordan}{Long
  et~al\mbox{.}}{2015}]%
        {ftran2}
\bibfield{author}{\bibinfo{person}{Mingsheng Long}, \bibinfo{person}{Yue Cao},
  \bibinfo{person}{Jianmin Wang}, {and} \bibinfo{person}{Michael Jordan}.}
  \bibinfo{year}{2015}\natexlab{}.
\newblock \showarticletitle{Learning Transferable Features with Deep Adaptation
  Networks}. In \bibinfo{booktitle}{\emph{Proceedings of the 32nd International
  Conference on Machine Learning}} \emph{(\bibinfo{series}{ICML '15})}.
  \bibinfo{pages}{97--105}.
\newblock


\bibitem[\protect\citeauthoryear{Ma, Kang, and Liu}{Ma et~al\mbox{.}}{2019a}]%
        {hgn19}
\bibfield{author}{\bibinfo{person}{Chen Ma}, \bibinfo{person}{Peng Kang}, {and}
  \bibinfo{person}{Xue Liu}.} \bibinfo{year}{2019}\natexlab{a}.
\newblock \showarticletitle{Hierarchical Gating Networks for Sequential
  Recommendation}. In \bibinfo{booktitle}{\emph{Proceedings of the 25th ACM
  SIGKDD International Conference on Knowledge Discovery \& Data Mining}}
  \emph{(\bibinfo{series}{KDD '19})}. \bibinfo{pages}{825--833}.
\newblock


\bibitem[\protect\citeauthoryear{Ma, Zhang, Cao, Jin, Wang, Liu, Ma, and
  Ren}{Ma et~al\mbox{.}}{2019b}]%
        {exrule19}
\bibfield{author}{\bibinfo{person}{Weizhi Ma}, \bibinfo{person}{Min Zhang},
  \bibinfo{person}{Yue Cao}, \bibinfo{person}{Woojeong Jin},
  \bibinfo{person}{Chenyang Wang}, \bibinfo{person}{Yiqun Liu},
  \bibinfo{person}{Shaoping Ma}, {and} \bibinfo{person}{Xiang Ren}.}
  \bibinfo{year}{2019}\natexlab{b}.
\newblock \showarticletitle{Jointly Learning Explainable Rules for
  Recommendation with Knowledge Graph}. In \bibinfo{booktitle}{\emph{The World
  Wide Web Conference}} \emph{(\bibinfo{series}{WWW '19})}.
  \bibinfo{pages}{1210--1221}.
\newblock


\bibitem[\protect\citeauthoryear{Mehrabi, Morstatter, Saxena, Lerman, and
  Galstyan}{Mehrabi et~al\mbox{.}}{2019}]%
        {fsurv19}
\bibfield{author}{\bibinfo{person}{Ninareh Mehrabi}, \bibinfo{person}{Fred
  Morstatter}, \bibinfo{person}{Nripsuta Saxena}, \bibinfo{person}{Kristina
  Lerman}, {and} \bibinfo{person}{Aram Galstyan}.}
  \bibinfo{year}{2019}\natexlab{}.
\newblock \bibinfo{title}{A Survey on Bias and Fairness in Machine Learning}.
\newblock
\newblock
\showeprint[arxiv]{1908.09635}~[cs.LG]


\bibitem[\protect\citeauthoryear{Nguyen, Hui, Harper, Terveen, and
  Konstan}{Nguyen et~al\mbox{.}}{2014}]%
        {fltbb14}
\bibfield{author}{\bibinfo{person}{Tien~T. Nguyen}, \bibinfo{person}{Pik-Mai
  Hui}, \bibinfo{person}{F.~Maxwell Harper}, \bibinfo{person}{Loren Terveen},
  {and} \bibinfo{person}{Joseph~A. Konstan}.} \bibinfo{year}{2014}\natexlab{}.
\newblock \showarticletitle{Exploring the Filter Bubble: The Effect of Using
  Recommender Systems on Content Diversity}. In
  \bibinfo{booktitle}{\emph{Proceedings of the 23rd International Conference on
  World Wide Web}} \emph{(\bibinfo{series}{WWW '14})}.
  \bibinfo{pages}{677--686}.
\newblock


\bibitem[\protect\citeauthoryear{Pariser}{Pariser}{2011}]%
        {fbori11}
\bibfield{author}{\bibinfo{person}{Eli Pariser}.}
  \bibinfo{year}{2011}\natexlab{}.
\newblock \bibinfo{booktitle}{\emph{The Filter Bubble: What The Internet Is
  Hiding From You}}.
\newblock \bibinfo{publisher}{Penguin Books Limited}.
\newblock


\bibitem[\protect\citeauthoryear{Rahman, Surma, Backes, and Zhang}{Rahman
  et~al\mbox{.}}{2019}]%
        {fwalk19}
\bibfield{author}{\bibinfo{person}{Tahleen Rahman}, \bibinfo{person}{Bartlomiej
  Surma}, \bibinfo{person}{Michael Backes}, {and} \bibinfo{person}{Yang
  Zhang}.} \bibinfo{year}{2019}\natexlab{}.
\newblock \showarticletitle{Fairwalk: Towards Fair Graph Embedding}. In
  \bibinfo{booktitle}{\emph{Proceedings of the Twenty-Eighth International
  Joint Conference on Artificial Intelligence, {IJCAI-19}}}.
  \bibinfo{pages}{3289--3295}.
\newblock


\bibitem[\protect\citeauthoryear{Rastegarpanah, Gummadi, and
  Crovella}{Rastegarpanah et~al\mbox{.}}{2019}]%
        {ffire19}
\bibfield{author}{\bibinfo{person}{Bashir Rastegarpanah},
  \bibinfo{person}{Krishna~P. Gummadi}, {and} \bibinfo{person}{Mark Crovella}.}
  \bibinfo{year}{2019}\natexlab{}.
\newblock \showarticletitle{Fighting Fire with Fire: Using Antidote Data to
  Improve Polarization and Fairness of Recommender Systems}. In
  \bibinfo{booktitle}{\emph{Proceedings of the Twelfth ACM International
  Conference on Web Search and Data Mining}} \emph{(\bibinfo{series}{WSDM
  '19})}. \bibinfo{pages}{231--239}.
\newblock


\bibitem[\protect\citeauthoryear{Rendle, Freudenthaler, Gantner, and
  Schmidt-Thieme}{Rendle et~al\mbox{.}}{2009}]%
        {bpr09}
\bibfield{author}{\bibinfo{person}{Steffen Rendle}, \bibinfo{person}{Christoph
  Freudenthaler}, \bibinfo{person}{Zeno Gantner}, {and} \bibinfo{person}{Lars
  Schmidt-Thieme}.} \bibinfo{year}{2009}\natexlab{}.
\newblock \showarticletitle{BPR: Bayesian Personalized Ranking from Implicit
  Feedback}. In \bibinfo{booktitle}{\emph{Proceedings of the Twenty-Fifth
  Conference on Uncertainty in Artificial Intelligence}}
  \emph{(\bibinfo{series}{UAI '09})}. \bibinfo{pages}{452--461}.
\newblock


\bibitem[\protect\citeauthoryear{Tang and Wang}{Tang and Wang}{2018}]%
        {caser18}
\bibfield{author}{\bibinfo{person}{Jiaxi Tang} {and} \bibinfo{person}{Ke
  Wang}.} \bibinfo{year}{2018}\natexlab{}.
\newblock \showarticletitle{Personalized Top-N Sequential Recommendation via
  Convolutional Sequence Embedding}. In \bibinfo{booktitle}{\emph{Proceedings
  of the Eleventh ACM International Conference on Web Search and Data Mining}}
  \emph{(\bibinfo{series}{WSDM '18})}. \bibinfo{pages}{565--573}.
\newblock


\bibitem[\protect\citeauthoryear{Togashi, Otani, and Satoh}{Togashi
  et~al\mbox{.}}{2021}]%
        {kgpl21}
\bibfield{author}{\bibinfo{person}{Riku Togashi}, \bibinfo{person}{Mayu Otani},
  {and} \bibinfo{person}{Shin'ichi Satoh}.} \bibinfo{year}{2021}\natexlab{}.
\newblock \showarticletitle{Alleviating Cold-Start Problems in Recommendation
  through Pseudo-Labelling over Knowledge Graph}. In
  \bibinfo{booktitle}{\emph{Proceedings of the 14th ACM International
  Conference on Web Search and Data Mining}} \emph{(\bibinfo{series}{WSDM
  '21})}. \bibinfo{pages}{931--939}.
\newblock


\bibitem[\protect\citeauthoryear{Veli{\v{c}}kovi{\'{c}}, Cucurull, Casanova,
  Romero, Li{\`{o}}, and Bengio}{Veli{\v{c}}kovi{\'{c}} et~al\mbox{.}}{2018}]%
        {gat18}
\bibfield{author}{\bibinfo{person}{Petar Veli{\v{c}}kovi{\'{c}}},
  \bibinfo{person}{Guillem Cucurull}, \bibinfo{person}{Arantxa Casanova},
  \bibinfo{person}{Adriana Romero}, \bibinfo{person}{Pietro Li{\`{o}}}, {and}
  \bibinfo{person}{Yoshua Bengio}.} \bibinfo{year}{2018}\natexlab{}.
\newblock \showarticletitle{Graph Attention Networks}. In
  \bibinfo{booktitle}{\emph{Proceedings of International Conference on Learning
  Representations}} \emph{(\bibinfo{series}{ICLR '18})}.
\newblock


\bibitem[\protect\citeauthoryear{Wang, Zhang, Wang, Zhao, Li, Xie, and
  Guo}{Wang et~al\mbox{.}}{2018}]%
        {rpnet18}
\bibfield{author}{\bibinfo{person}{Hongwei Wang}, \bibinfo{person}{Fuzheng
  Zhang}, \bibinfo{person}{Jialin Wang}, \bibinfo{person}{Miao Zhao},
  \bibinfo{person}{Wenjie Li}, \bibinfo{person}{Xing Xie}, {and}
  \bibinfo{person}{Minyi Guo}.} \bibinfo{year}{2018}\natexlab{}.
\newblock \showarticletitle{RippleNet: Propagating User Preferences on the
  Knowledge Graph for Recommender Systems}. In
  \bibinfo{booktitle}{\emph{Proceedings of the 27th ACM International
  Conference on Information and Knowledge Management}}
  \emph{(\bibinfo{series}{CIKM '18})}. \bibinfo{pages}{417--426}.
\newblock


\bibitem[\protect\citeauthoryear{Wang, Zhang, Zhao, Li, Xie, and Guo}{Wang
  et~al\mbox{.}}{2019d}]%
        {mkr19}
\bibfield{author}{\bibinfo{person}{Hongwei Wang}, \bibinfo{person}{Fuzheng
  Zhang}, \bibinfo{person}{Miao Zhao}, \bibinfo{person}{Wenjie Li},
  \bibinfo{person}{Xing Xie}, {and} \bibinfo{person}{Minyi Guo}.}
  \bibinfo{year}{2019}\natexlab{d}.
\newblock \showarticletitle{Multi-Task Feature Learning for Knowledge Graph
  Enhanced Recommendation}. In \bibinfo{booktitle}{\emph{The World Wide Web
  Conference}} \emph{(\bibinfo{series}{WWW '19})}. \bibinfo{pages}{2000--2010}.
\newblock


\bibitem[\protect\citeauthoryear{{Wang}, {Mao}, {Wang}, and {Guo}}{{Wang}
  et~al\mbox{.}}{2017}]%
        {kgesur17}
\bibfield{author}{\bibinfo{person}{Q. {Wang}}, \bibinfo{person}{Z. {Mao}},
  \bibinfo{person}{B. {Wang}}, {and} \bibinfo{person}{L. {Guo}}.}
  \bibinfo{year}{2017}\natexlab{}.
\newblock \showarticletitle{Knowledge Graph Embedding: A Survey of Approaches
  and Applications}.
\newblock \bibinfo{journal}{\emph{IEEE Transactions on Knowledge and Data
  Engineering}} \bibinfo{volume}{29}, \bibinfo{number}{12}
  (\bibinfo{year}{2017}), \bibinfo{pages}{2724--2743}.
\newblock


\bibitem[\protect\citeauthoryear{Wang, He, Cao, Liu, and Chua}{Wang
  et~al\mbox{.}}{2019a}]%
        {kgat19}
\bibfield{author}{\bibinfo{person}{Xiang Wang}, \bibinfo{person}{Xiangnan He},
  \bibinfo{person}{Yixin Cao}, \bibinfo{person}{Meng Liu}, {and}
  \bibinfo{person}{Tat-Seng Chua}.} \bibinfo{year}{2019}\natexlab{a}.
\newblock \showarticletitle{KGAT: Knowledge Graph Attention Network for
  Recommendation}. In \bibinfo{booktitle}{\emph{Proceedings of the 25th ACM
  SIGKDD International Conference on Knowledge Discovery \& Data Mining}}
  \emph{(\bibinfo{series}{KDD '19})}. \bibinfo{pages}{950--958}.
\newblock


\bibitem[\protect\citeauthoryear{Wang, He, Wang, Feng, and Chua}{Wang
  et~al\mbox{.}}{2019b}]%
        {ngcf19}
\bibfield{author}{\bibinfo{person}{Xiang Wang}, \bibinfo{person}{Xiangnan He},
  \bibinfo{person}{Meng Wang}, \bibinfo{person}{Fuli Feng}, {and}
  \bibinfo{person}{Tat-Seng Chua}.} \bibinfo{year}{2019}\natexlab{b}.
\newblock \showarticletitle{Neural Graph Collaborative Filtering}. In
  \bibinfo{booktitle}{\emph{Proceedings of the 42nd International ACM SIGIR
  Conference on Research and Development in Information Retrieval}}
  \emph{(\bibinfo{series}{SIGIR '19})}. \bibinfo{pages}{165--174}.
\newblock


\bibitem[\protect\citeauthoryear{Wang, Huang, Wang, Yuan, Liu, He, and
  Chua}{Wang et~al\mbox{.}}{2021}]%
        {kgin21}
\bibfield{author}{\bibinfo{person}{Xiang Wang}, \bibinfo{person}{Tinglin
  Huang}, \bibinfo{person}{Dingxian Wang}, \bibinfo{person}{Yancheng Yuan},
  \bibinfo{person}{Zhenguang Liu}, \bibinfo{person}{Xiangnan He}, {and}
  \bibinfo{person}{Tat-Seng Chua}.} \bibinfo{year}{2021}\natexlab{}.
\newblock \showarticletitle{Learning Intents behind Interactions with Knowledge
  Graph for Recommendation}. In \bibinfo{booktitle}{\emph{Proceedings of the
  Web Conference}} \emph{(\bibinfo{series}{WWW '21})}.
\newblock


\bibitem[\protect\citeauthoryear{Wang, Wang, Xu, He, Cao, and Chua}{Wang
  et~al\mbox{.}}{2019c}]%
        {kprn19}
\bibfield{author}{\bibinfo{person}{Xiang Wang}, \bibinfo{person}{Dingxian
  Wang}, \bibinfo{person}{Canran Xu}, \bibinfo{person}{Xiangnan He},
  \bibinfo{person}{Yixin Cao}, {and} \bibinfo{person}{Tat-Seng Chua}.}
  \bibinfo{year}{2019}\natexlab{c}.
\newblock \showarticletitle{Explainable Reasoning over Knowledge Graphs for
  Recommendation}. In \bibinfo{booktitle}{\emph{Proceedings of the
  Thirty-Fourth AAAI Conference on Artificial Intelligence, {AAAI-19}}}.
  \bibinfo{pages}{5329--5336}.
\newblock


\bibitem[\protect\citeauthoryear{Wang, Xu, He, Cao, Wang, and Chua}{Wang
  et~al\mbox{.}}{2020}]%
        {kgpmf20}
\bibfield{author}{\bibinfo{person}{Xiang Wang}, \bibinfo{person}{Yaokun Xu},
  \bibinfo{person}{Xiangnan He}, \bibinfo{person}{Yixin Cao},
  \bibinfo{person}{Meng Wang}, {and} \bibinfo{person}{Tat-Seng Chua}.}
  \bibinfo{year}{2020}\natexlab{}.
\newblock \showarticletitle{Reinforced Negative Sampling over Knowledge Graph
  for Recommendation}. In \bibinfo{booktitle}{\emph{Proceedings of The Web
  Conference 2020}} \emph{(\bibinfo{series}{WWW '20})}.
  \bibinfo{pages}{99--109}.
\newblock


\bibitem[\protect\citeauthoryear{Wu, Ahmed, Beutel, Smola, and Jing}{Wu
  et~al\mbox{.}}{2017}]%
        {rrn17}
\bibfield{author}{\bibinfo{person}{Chao-Yuan Wu}, \bibinfo{person}{Amr Ahmed},
  \bibinfo{person}{Alex Beutel}, \bibinfo{person}{Alexander~J. Smola}, {and}
  \bibinfo{person}{How Jing}.} \bibinfo{year}{2017}\natexlab{}.
\newblock \showarticletitle{Recurrent Recommender Networks}. In
  \bibinfo{booktitle}{\emph{Proceedings of the Tenth ACM International
  Conference on Web Search and Data Mining}} \emph{(\bibinfo{series}{WSDM
  '17})}. \bibinfo{pages}{495–503}.
\newblock


\bibitem[\protect\citeauthoryear{Yao and Huang}{Yao and Huang}{2017}]%
        {faircf17}
\bibfield{author}{\bibinfo{person}{Sirui Yao} {and} \bibinfo{person}{Bert
  Huang}.} \bibinfo{year}{2017}\natexlab{}.
\newblock \showarticletitle{Beyond Parity: Fairness Objectives for
  Collaborative Filtering}. In \bibinfo{booktitle}{\emph{Proceedings of the
  31st International Conference on Neural Information Processing Systems}}
  \emph{(\bibinfo{series}{NIPS '17})}. \bibinfo{pages}{2925--2934}.
\newblock


\bibitem[\protect\citeauthoryear{Yosinski, Clune, Bengio, and Lipson}{Yosinski
  et~al\mbox{.}}{2014}]%
        {ftran1}
\bibfield{author}{\bibinfo{person}{Jason Yosinski}, \bibinfo{person}{Jeff
  Clune}, \bibinfo{person}{Yoshua Bengio}, {and} \bibinfo{person}{Hod Lipson}.}
  \bibinfo{year}{2014}\natexlab{}.
\newblock \showarticletitle{How transferable are features in deep neural
  networks?}. In \bibinfo{booktitle}{\emph{Advances in Neural Information
  Processing Systems}} \emph{(\bibinfo{series}{NIPS '14},
  Vol.~\bibinfo{volume}{27})}. \bibinfo{pages}{3320--3328}.
\newblock


\bibitem[\protect\citeauthoryear{Yu, Zhang, Liang, and Zhang}{Yu
  et~al\mbox{.}}{2019}]%
        {marank19}
\bibfield{author}{\bibinfo{person}{Lu Yu}, \bibinfo{person}{Chuxu Zhang},
  \bibinfo{person}{Shangsong Liang}, {and} \bibinfo{person}{Xiangliang Zhang}.}
  \bibinfo{year}{2019}\natexlab{}.
\newblock \showarticletitle{Multi-Order Attentive Ranking Model for Sequential
  Recommendation}. In \bibinfo{booktitle}{\emph{Proceedings of the
  Thirty-Fourth AAAI Conference on Artificial Intelligence, {AAAI-19}}}.
  \bibinfo{pages}{5709--5716}.
\newblock


\bibitem[\protect\citeauthoryear{Yuan, Karatzoglou, Arapakis, Jose, and
  He}{Yuan et~al\mbox{.}}{2019}]%
        {nextitnet19}
\bibfield{author}{\bibinfo{person}{Fajie Yuan}, \bibinfo{person}{Alexandros
  Karatzoglou}, \bibinfo{person}{Ioannis Arapakis}, \bibinfo{person}{Joemon~M.
  Jose}, {and} \bibinfo{person}{Xiangnan He}.} \bibinfo{year}{2019}\natexlab{}.
\newblock \showarticletitle{A Simple Convolutional Generative Network for Next
  Item Recommendation}. In \bibinfo{booktitle}{\emph{Proceedings of the Twelfth
  ACM International Conference on Web Search and Data Mining}}
  \emph{(\bibinfo{series}{WSDM '19})}. \bibinfo{pages}{582--590}.
\newblock


\bibitem[\protect\citeauthoryear{Zhang, Yao, Sun, and Tay}{Zhang
  et~al\mbox{.}}{2019}]%
        {recsurv19}
\bibfield{author}{\bibinfo{person}{Shuai Zhang}, \bibinfo{person}{Lina Yao},
  \bibinfo{person}{Aixin Sun}, {and} \bibinfo{person}{Yi Tay}.}
  \bibinfo{year}{2019}\natexlab{}.
\newblock \showarticletitle{Deep Learning Based Recommender System: A Survey
  and New Perspectives}.
\newblock \bibinfo{journal}{\emph{ACM Comput. Surv.}} \bibinfo{volume}{52},
  \bibinfo{number}{1}, Article \bibinfo{articleno}{5} (\bibinfo{year}{2019}).
\newblock


\bibitem[\protect\citeauthoryear{Zhang, Humbert, Rahman, Li, Pang, and
  Backes}{Zhang et~al\mbox{.}}{2018}]%
        {tagv18}
\bibfield{author}{\bibinfo{person}{Yang Zhang}, \bibinfo{person}{Mathias
  Humbert}, \bibinfo{person}{Tahleen Rahman}, \bibinfo{person}{Cheng-Te Li},
  \bibinfo{person}{Jun Pang}, {and} \bibinfo{person}{Michael Backes}.}
  \bibinfo{year}{2018}\natexlab{}.
\newblock \showarticletitle{Tagvisor: A Privacy Advisor for Sharing Hashtags}.
  In \bibinfo{booktitle}{\emph{Proceedings of the 2018 World Wide Web
  Conference}} \emph{(\bibinfo{series}{WWW '18})}. \bibinfo{pages}{287--296}.
\newblock


\bibitem[\protect\citeauthoryear{Zheng, Liu, Li, and Wu}{Zheng
  et~al\mbox{.}}{2021}]%
        {cmeta21}
\bibfield{author}{\bibinfo{person}{Yujia Zheng}, \bibinfo{person}{Siyi Liu},
  \bibinfo{person}{Zekun Li}, {and} \bibinfo{person}{Shu Wu}.}
  \bibinfo{year}{2021}\natexlab{}.
\newblock \showarticletitle{Cold-start Sequential Recommendation via Meta
  Learner}. In \bibinfo{booktitle}{\emph{Proceedings of the Thirty-Sixth AAAI
  Conference on Artificial Intelligence}} \emph{(\bibinfo{series}{AAAI '21})}.
\newblock


\bibitem[\protect\citeauthoryear{Zhu, Hu, and Caverlee}{Zhu
  et~al\mbox{.}}{2018}]%
        {ften18}
\bibfield{author}{\bibinfo{person}{Ziwei Zhu}, \bibinfo{person}{Xia Hu}, {and}
  \bibinfo{person}{James Caverlee}.} \bibinfo{year}{2018}\natexlab{}.
\newblock \showarticletitle{Fairness-Aware Tensor-Based Recommendation}. In
  \bibinfo{booktitle}{\emph{Proceedings of the 27th ACM International
  Conference on Information and Knowledge Management}}
  \emph{(\bibinfo{series}{CIKM '18})}. \bibinfo{pages}{1153--1162}.
\newblock


\end{thebibliography}

\end{document}